\documentclass{bmcart}

\usepackage{amsthm,amsmath,amssymb}
\usepackage{caption}
\usepackage{subcaption}
\usepackage[pdftex]{graphicx}
\usepackage[utf8]{inputenc} 

\startlocaldefs
\def\E{\mathcal{E}} 
\def\F{\mathcal{F}} 
\def\R{\mathcal{R}} 
\def\Q{\mathcal{Q}}

 \newcommand{\PP}{\mbox{$\mathbb P$}}
\newcommand{\EE}{\mbox{$\mathbb E$}}
\def\c{{\rm cl}}
\def\supp{{\rm supp}}
\def\true{\texttt{true}}
\def\false{\texttt{false}}
\newtheorem{lemma}{Lemma}[section]
\newtheorem{theorem}{Theorem}

\newtheorem{corollary}[lemma]{Corollary}

\endlocaldefs

\begin{document}

\begin{frontmatter}

\begin{fmbox}
\dochead{Research}


\title{Autocatalytic sets in a partitioned biochemical network}


\author[
   addressref={aff1},                   
   email={joshiansmith@gmail.com}   
]{\inits{JI}\fnm{Joshua I} \snm{Smith}}
\author[
   addressref={aff1},
   email={mike.steel@canterbury.ac.nz}
]{\inits{MA}\fnm{Mike} \snm{Steel}}
\author[
   addressref={aff2},
         corref={aff1},   
   email={wim@SmartAnalytiX.com}
]{\inits{W}\fnm{Wim} \snm{Hordijk}}


\address[id=aff1]{
  \orgname{Biomathematics Research Centre, Department of Mathematics and Statistics, University of Canterbury}, 
  \city{Christchurch},                              
  \cny{NZ}                                    
}
\address[id=aff2]{%
  \orgname{SmartAnalytiX.com},
   \city{Lausanne},
  \cny{Switzerland}
}


\begin{artnotes}
\end{artnotes}

\end{fmbox}


\begin{abstractbox}

\begin{abstract} 
\parttitle{Background} 
In previous work, RAF theory has been developed as a tool for making theoretical progress on the origin of life question, providing insight into the structure and occurrence of self-sustaining and collectively autocatalytic sets within catalytic polymer networks. 
We present here an extension in which there are two ``independent" polymer sets, where catalysis occurs within and between the sets, but there are no reactions combining polymers from both sets. Such an extension reflects the interaction between nucleic acids and peptides observed in modern cells and proposed forms of early life.

\parttitle{Results} 
We present theoretical work and simulations which suggest that the occurrence of autocatalytic sets is robust to the partitioned structure of the network. We also show that autocatalytic sets remain likely even when the molecules in the system are not polymers, and a low level of inhibition is present. Finally, we present a kinetic extension which assigns a rate to each reaction in the system, and show that identifying autocatalytic sets within such a system is an NP-complete problem.

\parttitle{Conclusions} 
Recent experimental work has challenged the necessity of an RNA world by suggesting that peptide-nucleic acid interactions occurred early in chemical evolution. The present work indicates that such a peptide-RNA world could support the development of autocatalytic sets and is thus a feasible alternative worthy of investigation.

\end{abstract}


\begin{keyword}
\kwd{Origin of Life}
\kwd{Peptide-RNA World}
\kwd{Autocatalysis}
\end{keyword}


\end{abstractbox}
%

\end{frontmatter}




\section{Introduction}

Understanding the origin of life on Earth is an important and fascinating problem \cite{smi}. In order to shed light on the structure of early replicators and their mechanism of formation, various experimental approaches have been explored \cite{miller1953, hanc2003, vaidya2012, lili}. Due to the enormity of the task, experimental work alone seems unlikely to answer the question, and this has motivated several theoretical investigations \cite{eigen1977, kauf1986, raf2004, cot2007, dea13}. While one goal of theoretical work is to accelerate experimental progress (either in top-down construction of a minimal cell \cite{forster2006}, or the spontaneous formation of a self-replicating protocell from abiotic precursor molecules), links between theory and experiment have been scarce. Naturally, theoretical models are simplifications of real chemistry, and while such simplification enables progress, it may limit the conversation between theorists and experimentalists until the models more accurately reflect the complexity of real biochemical systems.

The combinatorial and stochastic aspects of theoretical work on the origin of life mean mathematics has an important role to play. The intuitive analogy between sets of reacting compounds and directed graphs was the motivation for Bollobas and Rasmussen's work on directed cycles in random graphs \cite{bollo1989}. In previous work \cite{raf2000, raf2004, raf2005, raf2011, raf2012a}, RAF theory has been developed as an effective tool for making progress on theoretical questions about the origin of life, based on initial work by Kauffman \cite{kauf1986, kauf1993}. In particular, it appears the emergence of collectively autocatalytic and self-sustaining sets of chemical reactions (RAF sets, defined later) is necessary for the origin of life to occur. Previous work has investigated the structure of such sets and the probability of their formation, leading to theoretical and empirical (simulation-based) results.

The general ideas behind RAF theory are not unique, and there are several related formalisms \cite{eigen1977, petri, cot2007}. However, in some cases, questions within the RAF framework have proven tractable while an equivalent question posed within an alternative formalism has not, perhaps because of the simplicity of the RAF model. On the other hand, it has been suggested that such simplicity limits our ability to draw conclusions about ``real" biochemical systems. However, the recent demonstration of the ability of RAF theory to link theoretical and experimental results \cite{vaidya2012, raf2012b}, together with the ongoing development of fresh theoretical ideas \cite{raf2013}, suggests that this framework continues to enable progress.

In this paper, we present a biologically relevant extension to the well-studied polymer model, formalising a network of molecules in which there are two ``independent" types of polymer, which are able to catalyse each others' (and their own) reactions, but cannot combine to form hybrid polymers. The motivation for this is the nature of the interaction between peptides and nucleic acids in the metabolic networks of modern cells. The importance of an extension addressing this mutually catalytic arrangement was highlighted in Kauffman's 1986 paper, in note (vii) (p. 14): {\em ``An independent then symbiotic coexistence of autocatalytic protein sets and template replicative polynucleotides would obviously be useful in prebiotic evolution.''} (While the present work does not address the templating ability of nucleic acids, this aspect has been studied previously \cite{raf2011,raf2012c}). Moreover, this extension is highly relevant in the light of recent experimental results from Li et al. \cite{lili}. In their paper, the authors propose that interactions between polypeptides and polynucleotides occurred very early in chemical evolution, providing an alternative to the hypothesis that life began in an RNA World \cite{rnaworld}. The authors state {\em ``The striking reciprocity of proteins and RNA in biology is consistent with our proposal: proteins exclusively catalyze nucleic acid synthesis; RNA catalyzes protein synthesis; and genetic messages are interpreted by the small ribosomal subunit, a ribonucleoprotein."} The reciprocity described here provides a clear motivation for theoretical investigation into the properties of these ``symbiotic" polymer systems.

We present theoretical results showing that RAF sets are just as likely to emerge in such systems as in those previously studied \cite{raf2011}, and it turns out that the result holds even for a more general system in which the molecules are not necessarily polymers, a small amount of inhibition is allowed, and the amount of catalysis varies freely across the reaction network. In previous work, catalysis has been assigned randomly with equal probability between each molecule and each reaction. The current work shows that RAF sets remain highly probable even under heterogenous catalysis, which is what we might expect to find in real biochemical networks.  

As a step toward increased chemical realism, we introduce the concept of a kinetic chemical reaction system, in which every reaction has an associated rate, and all molecules are lost via diffusion into the environment at a constant rate. We can in principle then search for RAFs in the system (as in previous work \cite{raf2004}) with the additional requirement that every molecule in the RAF must be produced at least as fast as it is used up or diffuses away - we call such an RAF a {\em kinetically viable RAF} (kRAF). 

\section{Definitions}
We will use the notation of Hordijk and Steel \cite{raf2004}. Consider a triple $(X, \R, F)$, where
\begin{itemize}
\item $X = \{ x_1, x_2, \dots \}$ is a (finite) set of {\em molecular species} or {\em molecule types};
\item $F \subset X$ is a distinguished subset of molecular species known as the {\em food set}, in which each species is assumed to be freely available in the environment;
\item $\R = \{ r_1, r_2, \dots \} $ is a (finite) set of chemically allowed {\em reactions};
\item Each reaction $r \in \R$ is an ordered pair $(A, B)$, where $A \subseteq X$ is a  multiset of {\em reactants} and $B \subseteq X$ is a multiset of {\em products}. We can represent a reaction as $a_1 + a_2 + \dots + a_n \rightarrow b_1 + b_2 + \dots b_m$. Note that the reactants $a_i$ are not necessarily distinct, and neither are the products $b_i$.
\end{itemize}
The triple $(X, \R, F)$ is therefore a set of molecular species together with the reactions that occur between them, intuitively visualised as a directed graph. For brevity, we will often use the term ``molecule" in place of ``molecular species" or ``molecule type". We also define $\rho(r)$ to be the set of all distinct reactants of the reaction $r$, and $\pi(r)$ to be the set of all distinct products of $r$. Then for any subset $\R'$ of $\R$, $\rho(\R') := \bigcup_{r \in \R'} \rho(r)$ and $\pi(\R') := \bigcup_{r \in \R'} \pi(r)$. Another useful concept will be the {\em support} of a reaction $r$, $\supp(r) := \rho(r) \cup \pi(r)$. Similarly, $\supp(\R') := \rho(\R') \cup \pi(\R')$ for any subset $\R'$ of $\R$. Informally, the support of a set of reactions is the set of all molecules consumed or produced by those reactions.

We can equip the triple $(X, \R, F)$ with a {\em catalysation assignment}  $C \subseteq X \times \R$, where $(x,r) \in C$ is understood to mean that the molecule $x$ {\em catalyses} reaction $r$: that is, $x$ accelerates $r$ but is unchanged by the reaction. A {\em chemical reaction system} (CRS) is now defined as a triple $(X, \R, F)$ together with a catalysation assignment $C$. We will denote a CRS $\Q$ by $\Q = (X, \R, F, C)$.   Fig. 1 shows an example of a CRS, within the {\em binary polymer model}, defined by Kauffman \cite{kauf1986} and well studied by Hordijk and Steel \cite{raf2004}. In this model, all molecule types are polymers over a 2-letter alphabet, and each reaction is either the ligation of two molecules into a longer polymer, or the cleavage of a single molecule into two shorter polymers.

The final important concept is that of the {\em closure} $\c_{\R'}(F)$ of the food set relative to a subset of reactions $\R' \subseteq \R$, formally defined as the minimal subset $W \subseteq X$ which contains $F$ and satisfies $\rho(r) \in W \Rightarrow \pi(r) \in W$ for all $r \in \R'$. Informally, $\c_{\R'}(F)$ is the set of all molecules that can be built up from the food set using only reactions in $\R'$.

Following \cite{raf2005}, we say that a  subset $\R'$ of $\R$ forms a {\em reflexively autocatalytic and food-generated set} (an RAF set) for $\Q$ provided that $\R'$ is non-empty and that:
\begin{itemize}
\item[(i)] All the reactants of each reaction in $\R'$ are contained in ${\rm cl}_{\R'}(F)$ (food-generated);
\item[(ii)] For each $r \in \R'$, there exists $(x,r) \in C$ such that $x \in \c_{\R'}(F)$ (reflexively autocatalytic).
\end{itemize}

We commonly use ``$F$-generated" in place of ``food-generated", and ``RAF" in place of ``RAF set". Informally, property (i) requires that the reactions in $\R'$ must be able to sustain themselves from the food set alone. Property (ii) requires that every reaction in $\R'$ must be catalysed, and furthermore that the catalysts must themselves be generated from the food set by that same set of reactions. 

These definitions are intended to capture properties of chemical networks that may have been important in the emergence of early replicators. Uncatalysed reactions in general proceed extremely slowly. We require catalysis so that molecules accumulate in concentrations sufficient to perform useful biochemical tasks. Otherwise, they would diffuse away before being able to play any role in the emergence of the first replicator.  Moreover, not only do catalysts greatly increase the reaction rates, they also lead to an equally dramatic  reduction in the variance of the rate of reactions ({\em c.f.} \cite{wol}, Fig. 6); this last feature would seem to be important for obtaining some degree of synchronicity in both early and present-day metabolism.   However, to allow the catalysts to come out of nowhere would be begging the question. So in addition, we require that the reactions generate their own catalysts from the food set. 

The idea of a set being $F$-generated requires that no molecules are needed before they have been produced. A set that fails to be $F$-generated could never have spontaneously built itself up from the food set, which is clearly a necessary condition for the development of early replicators from prebiotic chemistry.

Fig. 1 illustrates some ways in which a set can fail to be an RAF. The subset $\{r_1, r_2, r_3, r_5, r_7 \}$ fails to be reflexively autocatalytic (and so fails to be an RAF) since $r_7$ is uncatalysed. In the subset $\{r_1, r_2, r_3, r_5, r_6\}$ all reactions are catalysed, however the catalyst of $r_6$ is outside ${\rm cl}_{\{r_1, r_2, r_3, r_5, r_6\}}(F)$ (the reactions do not collectively generate all of their own catalysts), so this subset also fails to be reflexively autocatalytic. The subset $\{r_1, r_2, r_3, r_4, r_5\}$ is reflexively autocatalytic (since every reaction is catalysed, and all the catalysts are in $\c_{\{r_1,\dots,r_5\}}(F)$), but it is not $F$-generated, since the reactant $101$ of $r_4$ is not in the closure set (it cannot be created from the food set by the reactions $\{r_1,\dots,r_5\}$). However, the subset $\{ r_1, r_2, r_3, r_5\}$ is an RAF.   In fact, it is the largest RAF in the system, equal to the union of all RAFs in the system. Such an RAF is referred to as the maximal RAF subset or the {\em maxRAF}.

Given any catalytic reaction system $\Q = (X, \R, F, C)$, there is a fast (polynomial-time) algorithm which determines whether or not $\Q$ contains an RAF, and if so the algorithm constructs the maxRAF \cite{raf2004}.  We  use this algorithm in Section~\ref{sims}  to study the emergence of RAFs within simulations of the partitioned polymer system, defined in the following section.

\section{Partitioned polymer system}

All modern life utilises at least two polymers for linking information to structure and function:
nucleic acids (DNA/RNA) and peptides. Nucleic acids store and propagate genetic information, while peptides perform structural, catalytic and signalling roles {\em in vivo} in the form of proteins, enzymes and hormones. The interaction between peptides and nucleic acids is fundamental to the most important biochemical processes: peptides catalyse the replication of DNA and the synthesis of mRNA in transcription; at the ribosome, a combination of peptides and catalytic RNA molecules (ribozymes) catalyse the translation of mRNA, generating new peptide sequences. At the same time, each of these polymers catalyse reactions amongst themselves: for example, proteolytic enzymes catalyse the cleaveage of peptides, and a gene (DNA) could be considered to ``catalyse" transcription of mRNA by acting as a template (Fig. 2). Despite the mutual catalytic dependence of nucleic acids and peptides in living systems, these polymers are independent in the sense that there are no ``hybrid" polymers containing both nucleotide and amino acid monomers$^{a,b}$. In order to formalise these properties we introduce the following generalisation of the well studied polymer model \cite{raf2004}.

Consider a triple $(X, \R, F)$ within the polymer model. Let $X$, $\R$ and $F$ be partitioned as $X = \{ X_1,  X_2 \} $, $\R = \{ \R_1, \R_2 \}$ and $F=\{ F_1, F_2 \}$, where
\begin{itemize}
\item $X_1, X_2$ are {\em independent} sets of polymers;
\item $F_1 \subset X_1$ and $F_2 \subset X_2$ are independent food sets;
\item $\R_i$ is a set of ligation/cleavage reactions such that supp$(\R_i)\subseteq X_i$.
\end{itemize}
A {\em partitioned CRS} is now defined as a triple (partitioned as above) together with a catalysation assignment $C$. Note that (depending on the nature of $C$), molecules in either partition of $X$ can in principle catalyse or inhibit reactions in either partition of $\R$, but due to the condition that $\supp(\R_i) \subseteq X_i$, there can be no reactions involving molecules from both $X_1$ and $X_2$. We allows the partitions $X_1$ and $X_2$ to be sets of polymers over different sized monomer alphabets. For example, let the size of these alphabets be $k_1$ and $k_2$: then to model the interaction between a set of peptides ($X_1$) and a set of RNA polymers ($X_2$), set $k_1 = 20$, $k_2 = 4$. 

Fig. 3 shows a simple partitioned CRS within the binary polymer model. Previous work \cite{raf2004, raf2005} has demonstrated that in the standard, unpartitioned polymer model, RAFs are highly likely to be present in a CRS, given some mild requirements on the level of catalysis. Since the level of catalysis may vary across the network in the partitioned model, and since the partitioning makes the underlying structure of the reaction network qualitatively different, it is not obvious whether RAFs might be more or less likely to occur. This question is addressed more generally in the next section, where we prove a stronger result which is certainly sufficient to show that a partitioned CRS is just as likely to contain RAFs as an unpartitioned one. We will present the general result, before returning to the partitioned model.

\section{The probability of RAFs in general catalytic reaction systems}
\label{catraf}

It was shown in \cite{raf2005} that for a CRS within the polymer model, the level of catalysis (expected number of reactions catalysed per molecule) sufficient to produce RAF sets with a given probability increases linearly with $n$, the maximum length of polymers in the system. Here we extend this result to a general CRS in which the molecules are not necessarily polymers, and we invoke slightly weaker assumptions by allowing the catalysation rates to vary between reactions; we also allow for inhibition. Clearly, in such a setting, the concept of the length of a molecule has no meaning: instead we show that the average level of catalysis must increase linearly with $|\R| / |X|$, the ratio of the total number of reactions to the total number of molecules types.

In this generalised model we make two assumptions concerning catalysation:

\begin{itemize}
\item[(C1)] The events $\E(x,r)$ that molecule $x$ catalyses reaction $r$ are independent across all pairs $(x,r) \in X \times \R$.

\item[(C2)]  For some constant $K\geq 1$, the expected number of molecular species that catalyse any reaction is at most  $K$ times the expected number of molecular species that catalyse any other reaction.

\end{itemize}

Note that (C1) allows different molecule types to catalyse different numbers of reactions in expectation, since the probability that molecule type $x$ catalyses reaction $r$ can vary according to both $x$ and $r$ (in \cite{raf2005} it was assumed that the probability of $\E(x,r)$ depends only on $x$, not on $r$).

Before stating the main result of this section,  we require the following definition. We say that a triple $(X, \R, F)$ has a {\em species stratification} if and only if there is a nested sequence $\alpha_1 \subseteq \alpha_2 \subseteq \dots \subseteq \alpha_m = X$ such that $F = \alpha_t$ for some $t < m$ and the following hold, where  $X(1) := \alpha_1$ and $X(s) := \alpha_s - \alpha_{s-1}$ for $s \in \{ 2, \dots, m \}$.
\begin{itemize}
\item[(S1)]  The number of molecules in $\alpha_s$ grows no faster than geometrically with $s$. That is, $|X(s)| \leq k^{s}$ for some fixed $k\geq1$, for all $s \in \{ 1, \dots, m\}$;
\item[(S2)]  Every molecule in $\alpha_s$ that is not in $\alpha_{s-1}$ can be constructed from  molecules in $\alpha_{s-1}$ by a number of reactions that grows at least linearly with $s-1$.

More precisely,
for each $s \in \{t+1, \dots, m\}$, for all $x \in X(s)$, $\#\{r : x \in \pi(r),  \rho(r) \subseteq \alpha(s-1) \} \geq \nu(s-1)$ for some fixed $\nu\geq0$.
\end{itemize}

We now show that for any triple $(X,\R,F)$ the probability that $\Q = (X,\R, F, C)$ (where the random assignment $C$  satisfies  (C1) and (C2))  has an RAF (denoted $\PP(\exists \mbox{ RAF for } \Q)$) is, under certain conditions, determined by how the average catalysation rate compares to the simple ratio of the total number or reactions to the total number of molecules. 
The proof of part (a) of the following theorem is presented in the Appendix; part (b) follows immediately from a stronger result stated later (Theorem ~\ref{gentheorem2}) and the proof of that later result is also in the Appendix.

\begin{theorem}
\label{gentheorem}
For any  triple   $(X, \R, F)$ consisting of a set of molecular species, a set of reactions, and a subset of species (a `food set') consider the random CRS $\Q = (X, \R, F, C)$ formed by an assignment of  catalysation ($C$) under any stochastic process satisfying (C1) and (C2).

Let $\overline{\mu}$ be the average expected number of reactions that are catalysed by a molecular species (averaged over all molecular species in $X$).

\begin{itemize}
\item[(a)] If $\overline{\mu} \leq \lambda \cdot \frac{|R|}{|X|}$ then the probability that there exists an RAF for $\Q$ is at most $\phi(\lambda)$, where
 $\phi(\lambda)=1 - (1-\lambda/K)^{\frac{\tau}{2}\left(|F|^2+|F|\right)}  \rightarrow 0$ as $\lambda \rightarrow 0$ ($\tau$ is a constant described in the proof). 

\item[(b)] Suppose that   $(X,\R,F)$ has a species stratification.  If $\overline{\mu} \geq \lambda \cdot \frac{|R|}{|X|}$ then the probability that there exists an RAF for $\Q$ is at least $1 - \psi(\lambda),$  where $\psi(\lambda) = \frac{k(ke^{-\nu\lambda})^t}{1 - ke^{-\mu\lambda}}  \rightarrow 0$ exponentially fast as $\lambda \rightarrow \infty$.
\end{itemize}
\end{theorem}

The results in Section~\ref{partitionedCRS} show that as the level of catalysis is increased past some threshold there is a transition in the probability of the existence of RAFs. This is to be expected as it is well known in combinatorics that every monotone increasing property of subsets of a set has an associated threshold function \cite{bollobas2}. Consideration of the definitions of reflexively autocatalytic and $F$-generated reveals that the RAF property is monotone on the subsets of the set of possible catalysis arcs from molecules to reactions in a CRS, so the RAF property has a threshold function. Theorem \ref{gentheorem} tells us the form of this function (namely $|R|/|X|$).

\subsection{Remarks}
\begin{itemize}
\item The proof of part (b) involves the construction of a RAF involving every molecule in $X$ (that is, $\supp(\R') = X$). However, in general, this RAF will involve only a subset of the reactions in $\R$.

\item In general, the definition of a species stratification seems rather artificial: while a CRS within the simple (unpartitioned) polymer model naturally admits a species stratification (since we just let $\alpha_s$ be the set of all polymers up to length $s$), it would be a non-trivial exercise to find a species stratification for a CRS with molecules that are not polymers.  Nevertheless, Theorem~\ref{gentheorem} shows that the molecules in a CRS being polymers is sufficient but not necessary, and we will see shortly that in the partitioned polymer model a species stratification also applies.

\end{itemize}

\section{The probability of RAFs in a partitioned CRS} \label{partitionedCRS}

In light of Theorem~\ref{gentheorem}, in order to show that the same linear catalysis requirement that applies for an unpartitioned CRS holds for a partitioned one, we need only show that a partitioned CRS has a species stratification, and construct a set $C$ satisfying (C1), (C2). In previous work \cite{raf2013}, $C$ was often generated by randomly assigning catalysis as follows: let each element of $X \times \R$ be included in $C$ with some fixed probability $p$. When studying metabolic network data from real organisms, we might expect to find that this model does not match the observed pattern of catalysis: for example, it might be the case that peptides tend to catalyse more reactions involving other peptides than reactions involving nucleic acids. To allow for this possibility in a partitioned CRS, we allow the likelihood of catalysis to vary depending on both the nature of the catalyst and the nature of the molecules involved in the reaction. Specifically, we define the matrix ${\bf P}$ where, for any molecule $x \in X_i$ and any reaction $r \in \R_j$, the probability $p(x,r)$ that $x$ catalyses $r$ is given by the $ij$th entry of ${\bf P}$. For example, in a CRS generated using the matrix
$${\bf P} =
10^{-6}\cdot\begin{bmatrix}
10 & 1 \\
2 & 10 \\
\end{bmatrix}$$
we would expect to observe around ten times more catalysis within partitions than between them, and twice as much catalysis of reactions in $\R_1$ by molecules in $X_2$ than of reactions in $\R_2$ by molecules in $X_1$. 

In what follows, consider a partitioned CRS $\Q=(X,\R,C,F)$ which is {\em complete}: that is, both $X_1$ and $X_2$ contain every possible polymer up to length $n_1$ and $n_2$ (respectively), and $\R_1$ (respectively $\R_2$) contains every possible reaction between the molecules in $X_1$ (respectively $X_2$). Let $F_1$ (respectively $F_2$) be all the molecules in $X_1$ (respectively $X_2$) up to some length $t < \min\{ n_1, n_2 \}$. Finally, for a molecule $x \in X$, let $|x|$ denote the length of $x$ (i.e. the number of monomer units in $x$).

For $X_1$, define the stratification
$$\alpha_1 \subseteq \alpha_2 \subseteq \dots \subseteq \alpha_t \subseteq \dots \subseteq \alpha_{n_1} = X_1$$
where $\alpha_s$ consists of all the molecules in $X_1$ such that $1 \leq |x| \leq s$. It will prove useful to define $X_1(1) := \alpha_1$ and for $s \in \{2 , \dots, n_1 \}$, $X_1(s) := \alpha_s - \alpha_{s-1}$. Similarly for $X_2$, define the stratification
$$\beta_1 \subseteq \beta_2 \subseteq \dots \subseteq \beta_t \subseteq \dots \subseteq \beta_{n_2} = X_2$$
and let $X_2(s)$ be defined similarly to $X_1(s)$. Note that $|X_i(s)| = k_i^s$. Defining $n_{\min} := \min \{ n_1, n_2 \}$ and $n_{\max} := \max \{ n_1, n_2 \}$, these stratifications are combined into a single stratification of the set $X$ as follows:
\begin{itemize}
\item for $1 \leq s \leq n_{\min}$, $\gamma_s := \alpha_s \cup \beta_s$;
\item for $n_{\min} < s \leq n_{\max}$, $$\gamma_s := 
\begin{cases}
\alpha_s, & \mbox{if } n_1 > n_2 \\
\beta_s, & \mbox{if } n_2 > n_1 
\end{cases}.$$
\end{itemize}
Note that $F = \gamma_t$. Now define $X(1) := \gamma(1)$ and for $s \in \{2, \dots, n_{\max} \}, X(s) := \gamma(s) - \gamma(s-1)$, and consider the size of each set $X(s)$. The maximum size of $X(s)$ over all $s$ is $k_1^s + k_2^s$, hence $(k_1 + k_2)^s$ is strictly greater than $|X(s)|$ for all $s \in \{1, \dots, m \}$, so the partitioned CRS satisfies (S1). To see that it also satisfies (S2), we need only note that for any molecule type $x \in X(s)$ where $s \in \{t+1, \dots, m\}$, $|x| = s$, so there are a maximum of $s-1$ ways $x$ could be constructed from shorter molecule types (i.e. molecule types in $\gamma_{s-1}$). Since $\R_1$ and $\R_2$ are both complete, every such reaction exists and there are in fact precisely $s-1$ reactions generating $x$ from $\gamma_{s-1}$, so take $\nu=1$. We conclude that the complete partitioned CRS has a species stratification.

It remains to show that the catalysation assignment $C$ described above satisfies (C1), (C2). For each pair $(x,r) \in X \times \R$, the probability that $x$ catalyses $r$ is dependent only on which partitions $x$ and $r$ are from, so (C1) clearly holds. The following expression gives the expected number of species that catalyse any given reaction:
$${\bf P}^T 
\begin{bmatrix}
|X_1| \\
|X_2| \\
\end{bmatrix} = 
\begin{bmatrix}
p_{11} & p_{21} \\
p_{12} & p_{22} \\
\end{bmatrix}
\begin{bmatrix}
|X_1| \\
|X_2| \\
\end{bmatrix}
= 
\begin{bmatrix}
c_1 \\
c_2 \\
\end{bmatrix},$$
where $c_i$ is the expected number of species in $X$ that catalyse any given reaction in $\R_i$. Noting that $p_{ij} \in \left[0,1\right]$ and that 
$$|X| = |X_1| + |X_2| = \sum_{s=1}^{n_1} k_1 ^ s + \sum_{s=1}^{n_2} k_2 ^ s,$$
clearly $c_1, c_2$ are finite. Hence taking $K = \max \{ c_1/c_2, c_2/c_1\}$ shows that (C2) holds also. We conclude that Theorem~\ref{gentheorem} applies to a partitioned CRS.

\section{Simulations of partitioned chemical reactions systems}
\label{sims}

Previous simulations of chemical reaction systems \cite{raf2004, raf2011} have focussed on those which are {\em complete} ($X$ contains every molecule up to some maximum length $n$, and $\R$ contains every possible cleavage/ligation reaction between the molecules of $X$) and those in which catalysis is assigned randomly such that every molecule has the same fixed probability of catalysing any reaction. In \cite{raf2005,raf2011}, it was shown both theoretically and computationally that in a `classic' CRS with only one partition, the level of catalysis (expected number of reactions catalysed per molecule) sufficient to generate RAFs with a given probability (e.g. 0.5) increases linearly with $n$. Furthermore, simulations show that the linear relationship is not steep: when $n=10$, the required level of catalysis is around $1.29$, and when $n=20$, the required level of catalysis increases only to $1.48$ \cite{raf2011}. Since this level of catalysis is chemically realistic \cite{moonlight}, the results suggest that RAFs may be quite probable in real biochemical polymer networks. 

Theorem~\ref{gentheorem} assures us that the linear increase in the required level of catalysis seen in the original model also applies to the partitioned model. However, it is not obvious whether or not the same realistic level of catalysis will be seen in the latter. In particular, because the partitioned model is highly flexible in terms of possible patterns of catalysis, it is interesting to ask how the pattern of catalysis affects the probability of RAF formation. In order to address this question, we simulated a partitioned CRS in which each partition is complete, with $k_1=k_2=2$, $n=10$, and the food set consists of all monomers and dimers. This CRS was simulated under three different catalytic assignments (Fig. 4). In order to isolate the effect of the pattern of catalysis on the probability of RAF formation, the level of catalysis is constant across all three scenarios for any given value of $p$. We generated 500 instances of each model at a range of values of $p$ (corresponding to a range of levels of catalysis) and searched for RAFs using the algorithm from \cite{raf2004}.  Analyses were performed on an IBM Power755 cluster comprising 13 nodes, each with 32 CPUs  running Linux 11.1 (a total of 416 CPUs).

Fig. 5 shows, at each level of catalysis, the fraction of the 500 instances which were found to contain an RAF, for each of the three models investigated. All three models display a sharp transition in the probability of RAF formation as the level of catalysis increases, familiar from simulations of classic CRSs \cite{raf2004}. The uniform and reciprocal models display extremely similar results. The level of catalysis required to give 50\% probability of RAF formation in the non-selfish models ($\approx1.3$) is slightly higher than that in the selfish model ($\approx1.25$), indicating that during this transition, RAFs are slightly more likely in a selfish CRS than a CRS with some catalysis between partitions. When the level of catalysis is 1.29, around 75\% of instances of the selfish CRS contain an RAF, which is to be expected: the same level of catalysis in a single partition CRS with $n=10$ gives 50\% probability of RAF formation, and since here the selfish CRS essentially consists of two independent copies of the single partition CRS, the probability of finding an RAF is $1 -(1-0.5)^2= 0.75$. 

Fig. 5 also shows that, as the level of catalysis is increased past the transition level, the fraction of CRSs containing an RAF in the uniform and reciprocal models approaches 100\% slower than in the selfish model. However, by the time the catalysis level has reached $1.7$, all three models produce RAFs close to 100\% of the time. These results indicate that the pattern of emergence of RAFs in partitioned chemical reaction systems is very similar to that in simple systems. Moreover, it is clear that even under widely varying patterns of catalysis, partitioned systems develop RAFs with high probability. 

It is interesting to note that the property distinguishing the selfish model from the uniform and reciprocal models (whose results are identical, but differ from those of the selfish model) is the absence of between-partition catalysis. This binary contrast is present even though the uniform and reciprocal models differ greatly from each other in their patterns of catalysis: there is twice as much between-partition catalysis in the reciprocal model as there is in the uniform model, and the reciprocal model contains no within-partition catalysis. 

Despite the above similarity in the probability of emergence of RAFs between all three models, Fig. 6 shows that the dependence of the size of the maxRAF on the level of catalysis is qualitatively different depending on the pattern of catalysis. At low catalysis levels, all three models tend to contain only RAFs consisting of a single reaction, catalysed by one of its own reactants or products. At the threshold level of catalysis at which all 3 models begin to develop RAFs with higher probability, the number of reactions contained in the maxRAF in the selfish model increases faster than in the non-selfish models (which again display very similar results). However, after a short delay, the number of reactions in the non-selfish models rapidly increases, matching the equivalent value in the selfish model and then exceeding it. As the level of catalysis is increased further past the transition point, the rate of growth in the non-selfish models gradually decreases again, and all 3 models appear to converge on the same values of the average maxRAF size. This asymptotic behaviour makes sense: at higher levels of catalysis, the partition to which any particular catalyst belongs has less bearing on whether or not the reaction in question is part of an RAF set. 
Fig. 6 also shows how the average number of molecules contained in the maxRAF (expressed as a proportion of all the molecules in $X$) depends on the level of catalysis (more formally this is $|\c_{\R'}(F)|/|X|$, where $\R'$ is the maxRAF). The pattern of growth is similar to that seen in the number of reactions. However, one important contrast is that, while the maxRAF quickly grows to contain the majority of the molecules in $X$, at a given level of catalysis it contains only a relatively small proportion of the reactions in $\R$. Thus as the level of catalysis is increased beyond that shown in Fig. 6, we should expect the average proportion of molecules in the maxRAF to quickly approach 1.0, while the average proportion of reactions in the maxRAF continues to increase linearly. Not until a much higher level of catalysis will the maxRAF contain 100\% of the reactions in the system.

Overall, the above results show that when $n=10$, a partitioned CRS behaves very similarly to a classic CRS in terms of RAF emergence. In order to address the question of whether this is true for general values of $n$, we repeated the experiments at $n=15$ (Figs. 7 and 8). For each of the three models, the level of catalysis required to attain a given probability of RAF formation is higher, which is to be expected given previous theoretical and experimental work on the original model. However, while the selfish model undergoes a sharp transition similar to the $n=10$ case, both the uniform and reciprocal models undergo a more gradual increase in the probability of RAF formation as the level of catalysis increases from around $1.3$ up to $2.0$ (Fig. 7). Once again, the latter two models exhibit almost identical results, which is surprising given the difference in their pattern of catalysis. The level of catalysis required to give a 50\% probability of RAF formation in the reciprocal and uniform models has increased from around $1.3$ ($n=10$) to $1.45$ ($n=15$), while the increase in the same figure for the selfish model is smaller, going from around $1.25$ to around $1.32$. However, the increased level of catalysis in the non-selfish models remains chemically realistic. Fig. 7 also suggests that as the level of catalysis is further increased, the fraction of CRSs containing an RAF for the non-selfish models will approach $1$ monotonically, as observed for $n=10$ (Fig. 5).

Fig. 8 shows how the average size of the maxRAF (in terms of number of reactions, and number of molecules) changes as the level of catalysis is increased. Whereas for $n=10$ the maxRAF initially grew most quickly for the selfish model, these plots do not show the same early growth spurt: instead, all models appear to begin the transition at around the same level of catalysis. It is possible that the resolution was not high enough to detect the phenomenon: simulating smaller increments in $p$ and a greater number of instances around this transition zone may reveal that it still occurs at $n=15$. Other than this, the plots are similar to those in Fig. 6. RAF sets in the non-selfish models grow faster both in number of reactions and number of molecules, and not until a level of catalysis around $2.0$ does the selfish model catch up. This is much later than in the $n=10$ case, which is particularly interesting given that at this level of catalysis the selfish model is developing RAFs with higher probability than the non-selfish models (Fig. 7).

\subsection{Discussion}
We chose here to investigate only the cases when $n=10$ and $n=15$, since computational constraints limit the feasibility of repeating the experiments for more and/or larger values of $n$. However, inferences can be made about other values of $n$, especially in the light of Theorem~\ref{gentheorem}, which shows that a linear increase (with $n$) in the level of catalysis is necessary and sufficient to maintain RAFs with a given probability in a partitioned CRS. After producing similar results to the above for further values of $n$, it would be interesting to use least squares regression to explicitly express the linear dependence (on $n$) of the level of catalysis required to give 50\% probability of RAF formation for various patterns of catalysis, and compare these with the linear formulae produced in \cite{raf2011} for the original model. Based of Figs. 5 and 7, we expect to see a steeper relationship for the non-selfish (uniform and reciprocal) models than for the selfish model.

While all three models begin to develop RAFs with high probability above the threshold level of catalysis, it is clear that the selfish model develops RAFs somewhat more reliably (with higher probability at lower catalysis levels) than the non-selfish models. Furthermore, the difference between the selfish and non-selfish models is more apparent at $n=15$ than $n=10$, and in the light of the result of Theorem~\ref{gentheorem}, the difference looks likely to become more marked as $n$ increases. On the other hand, as pointed out by philosopher Roger White \cite{white}, the probability of a mechanism proposed to play a role in the origin of life may not be a sound metric by which to judge the validity of that mechanism (Elliott Sober makes a related argument in response to Richard Dawkins in \cite{sob} pp.50-51).  In terms of RAF theory, this means that the probability of RAF formation might not be the best way to decide which models have the most potential to shed light on the origin of life question.

However, the results show another difference between the selfish and non-selfish models that is worth noting. Figs 6 and 8 both suggest that the size of RAF sets in the selfish model is significantly lower in the selfish model than in the non-selfish models (excluding the brief window immediately around the threshold level of catalysis in which RAFs in the selfish model grow faster at $n=10$). This larger size of RAF sets in the non-selfish models is worth noting: since RAF sets can often be decomposed into constituent RAFs (subRAFs), larger RAFs are likely to contain more of these autocatalytic subsets. It was suggested in \cite{raf2012d, vasa12} that this modular structure might be important for the potential evolution of RAF sets. Specifically, the ability of RAF sets to gain and lose subRAFs might be a mechanism by which RAF sets can evolve and compete with each other, a process which might favour characteristic combinations of subRAFs, in a primitive form of selection. This transition from a purely self-replicating set of molecules to a complex autocatalytic set which replicates imperfectly while remaining robust to changes in the environment is essential, if RAF sets are to give rise to a replicator capable of gradual, open-ended Darwinian evolution.

We have investigated three different patterns of catalysis. Due to the inherent flexibility of the partitioned model, there are various other qualitatively different patterns that could be explored. In each of the above systems, the catalysis matrix ${\bf P}$ is symmetric. Even with this restriction in place, there is a continuum between totally reciprocal and totally selfish catalysis, and we examined only the middle point and the two extremes of that continuum here. We expect to observe a similar pattern of RAF emergence in other systems, where both intra- and inter-partition catalysis occur, but not in equal amounts. Based on Figs. 5 and 7, if we were to begin with a selfish system and gradually increase the level of inter-partition catalysis, we should expect to see a shift in the pattern of RAF development, becoming more like the uniform and reciprocal models examined here. This change should be complete by the time the catalysis becomes uniform, so must occur somewhere between `selfish' and `uniform'. It would be interesting to determine at what point this transition occurs, and how sharp it is. A further extension would be to investigate systems in which ${\bf P}$ is not symmetric: for example, where one partition dominates as a source of catalysts for the system. Given the main motivation behind this investigation, and the observation that peptides appear to be far more catalytically active than nucleic acids \cite{moonlight}, this particular extension seems highly relevant.

Based on structural complementarity between polypeptide and RNA helices \cite{cart74} and more recent experimental work demonstrating high catalytic proficiency of ancestrally related primitive forms of enzymes involved in translation \cite{pha07,li11,lili}, Carter and colleagues have suggested that the interactions between polypeptides and RNA may have played a key role in early chemical evolution in a ``peptide-RNA world". Our theoretical results show that a system with two different types of polymer with reciprocity of function similar to that of proteins and RNA, produces autocatalytic sets at similarly realistic levels of catalysis to one composed of a single type of polymer (such as an RNA-world or system of peptides). Therefore, the results presented here suggest that the alternative scenario proposed by Carter and colleagues is feasible.

\section{Extensions: Closure, Inhibition and Reaction Rates}

The current definition of an RAF is limited because it ignores inhibition and reaction rates. The latter is problematic because those reactions generating required reactants which proceed too slowly, or those which use up required reactants and proceed too fast, may prevent an RAF set from persisting in a dynamic environment. While the lack of inhibition and kinetics may be seen as a severe restriction, it is useful because it allows us to compute RAFs in polynomial time. These RAFs could then be examined to test if they are viable given known inhibition or reaction rate data.

Alternatively, we could build this into the definition of a stronger type of RAF and ask if there is an efficient algorithm to find them. In this section we explore the latter approach.  We  consider RAFs that are viable under reaction rates and show that determining whether or not they exist in an arbitrary catalytic reaction system  turns out to be NP-complete (the same NP-completeness was earlier shown for inhibition, and we use this result to establish the NP-completeness for the reaction rates problem).

Consideration of these
factors (inhibition and reaction rates)  requires distinguishing between RAFs that are `closed' and those that are not (this distinction is not important in the absence of inhibition and dynamics).  Thus we first introduce and discuss this property, before considering the definition and properties of RAFs that allow inhibition or reaction rates.

\subsection{Closed RAFs}

Given a CRS $\Q = (X, \R, C, F)$, a subset $\R'$ of $\R$ is a {\em closed RAF} if and only if the following conditions hold:
\begin{itemize}
\item[1)] $\R'$ is an RAF;
\item[2)] for every $r \in \R$ for which there is a pair $(x,r) \in C$ such that $\{x\} \cup \rho(r) \subseteq {\rm cl}_{\R'}(F)$,  $r \in \R'$.
\end{itemize}

Informally, a closed RAF captures the idea that ``any reaction that can occur, will occur". If all the reactants and at least one catalyst of a reaction $r \in \R$ are generated by the reactions in $\R'$, then it seems reasonable to expect that the reaction $r$ will occur, and so we should expect that $r$ is included in $\R'$. If $r$ is not included, then it is natural to consider adding it to $\R'$, in order that the extended set $\R' \cup \{r\}$ comes closer to containing all the reactions for which it generates all the necessary molecules. In order to formalise this notion, we introduce the idea of the {\em closure} of an RAF, defined as the smallest closed RAF which contains the RAF. Given an RAF $\R'$, we can construct its closure $\overline {\R'}$ as follows: let $\R' = K_0$, and let $K_{i+1} = K_i \cup L_i$, where $L_i$ is the set of all $r \in \R \setminus K_i$ such that there exists a pair $(x,r) \in C$ and $\{x\} \cup \rho(r) \subseteq \c_{K_i}(F)$. Then, $\overline {\R'}$ is the final set $K_n$ in the sequence of nested sets $\R' = K_0 \subset K_1 \subset K_2 \subset \dots \subset K_n$, where $n$ is the first value of $i$ for which $K_i = K_{i+1}$.

Note that an RAF $\R'$ is a closed RAF if and only if $\R' = \overline {\R'}$. Note also that while the union of two RAFs is also an RAF, the union of two closed RAFs is not necessarily a closed RAF (though it is an RAF).

One notable property of closed RAFs is that, unlike RAFs that are not closed, we can reconstruct the network of reactions given only a ``list" of the molecules involved in the network, as follows. 
\begin{lemma}
A closed RAF $\R' \subseteq \R$ is determined entirely by the subset of molecules $F \cup {\rm supp}(\R')$ and the CRS $\Q= (X, \R, C, F)$.
\label{closedlemma}
\end{lemma}
{\bf Proof.} Consider the set $\R^*$ reconstructed from $F\cup \supp(\R')$ as follows:
\begin{itemize}
\item[1)] Add  to $\R^*$ every reaction $r \in \R$ for which ${\rm supp}(r) \subseteq F \cup {\rm supp}(\R')$.
\item[2)] Remove from $\R^*$ any reaction $r$ for which there does not exist an $x \in F\cup {\rm supp}(\R')$ such that $(x,r) \in C$.
\end{itemize}

We will show that $\R^* = \R'$ by establishing the set inclusions $\R' \subseteq \R^*$ and  $\R^* \subseteq \R'$. First, consider some $r \in \R'$. Clearly $\supp(r) \subseteq \supp(\R')$. It remains to show that there exists some $x \in F \cup \supp(\R')$ such that $(x,r) \in C$. Since $\R'$ is an RAF for $(\Q,F)$, by Lemma 4.3 of \cite{raf2004}, $\c_{\R'}(F) = F \cup \pi(\R')$, which together with the definition of $F$-generated and of the support implies that 
 \begin{equation}
 F\cup \supp(\R') = \c_{\R'}(F).
\label{supp}
\end{equation}
By definition of reflexively autocatalytic, it follows from \eqref{supp} that for all $r \in \R'$, there exists $x \in F \cup \supp(\R')$ such that $(x,r)\in C$. Therefore every reaction in $\R'$ fits the criteria for inclusion in $\R^*$, and we conclude that $\R' \subseteq \R^*$.
 
Next consider some $r \in \R^*$. Then by the rules of construction of $\R^*$, $\supp(r) \subseteq F\cup \supp(\R')$ and there exists an $x \in F \cup \supp(\R')$ such that $(x,r) \in C$.  By (\ref{supp}), such an $x$ is in $\c_{\R'}(F)$, and also by (\ref{supp}), $\supp(r) \subseteq \c_{\R'}(F)$ so certainly $\rho(r) \subseteq \c_{\R'}(F)$. Then since $\R'$ is a closed RAF, $r \in \R'$ by definition. We conclude that $\R^* \subseteq \R'$, which together with the previous result proves that $\R^*=\R'$. $\hfill\blacksquare$

\begin{corollary}
If $\R'$ is an RAF, then given only $F \cup \supp(\R')$ and the CRS $\Q = (X, \R, C, F)$, we can construct its closure $\overline {\R'}$.
\end{corollary}
{\bf Proof.} If $\R'$ is a closed RAF, then $\R' = \overline {\R'}$ so the assertion holds trivially by the previous lemma. 

Hence suppose $\R'$ is not closed. Then there is at least one reaction $r^* \in \R \setminus \R'$ such that there exists a pair $(x,r^*)\in C$ and $\{x\} \cup \rho(r^*) \in \c_{\R'}(F)$. Construct the set of reactions $\R^*$ from $F \cup \supp(\R')$ (as in Lemma~\ref{closedlemma}). Since we did not use the fact that the RAF was closed in the first part of the proof of the lemma, we can apply the same argument to see that $\R' \subseteq \R^*$. 

Now consider some $r \in \R^*$. Then by the rules of construction of $\R^*$, $\supp(r) \subseteq F \cup \supp(\R')$, and there exists some $x \in F \cup \supp(\R')$ such that $(x,r) \in C$. Then by Equation \eqref{supp} in the proof of Lemma~\ref{closedlemma} (again, this applies since we did not assume the RAF was closed in that part of the proof), $\R^*$ contains every $r^* \in \R \setminus \R'$ such that there exists a pair $(x,r) \in C$ and  $\{x\} \cup \rho(r^*) \in \c_{\R'}(F)$. At this point, we identify $\R'$ with the set $K_0$ and $\R^*$ with the set $K_1 = K_0 \cup L_0$ described in the preamble to Lemma~\ref{closedlemma}. We can then follow the same process described in the preamble, constructing a sequence of nested sets $\R' = K_0 \subset \dots \subset K_n$, where $K_n$ is by definition equal to $\overline {\R'}$. $\hfill\blacksquare$

\subsection{Inhibition}

In order to discuss the impact of molecules inhibiting reactions, we begin with the following definitions.

Given a CRS $\Q=(X, \R, F,C)$ an {\em inhibition assignment} is a subset $I$ of $X\times \R$ where $(x,r) \in I$ means that molecular species $x$ {\em inhibits} reaction $r$.
We say that a subset $\R'$ of $\R$ is an $I$-viable RAF for $\Q$ if and only if all of the following hold:
\begin{itemize}
\item[(a)] $\R'$ is an RAF for $\Q$;
\item[(b)] $\R'$ is closed;
\item[(c)] No reaction in $\R'$ is inhibited by any molecule in ${\rm cl}_{\R'}(F)$.
\end{itemize}

The motivation for insisting that $\R'$ be closed is as follows: Suppose that $\R'$ involves a reaction that is inhibited by some product $x'$ of a reaction $r'$ that is not in $\R'$.
Now if the reactants, and at least one catalyst of $r'$ are present as products of reactions in $\R'$ (or elements of $F$) then there is no reason for $r'$ not to proceed and for $x'$ not to be produced. In that case  $\R' \cup \{r'\}$, and any set containing it, would no longer be an RAF.

The concept of an RAF subject to inhibition was formalized and studied briefly in  \cite{raf2005}, but there condition (b) was not imposed.  This paper established that the problem of determining whether or not a CRS contains an RAF that is $I$-viable for $\Q$ is an NP-complete problem.  It is pertinent therefore to ask whether the  addition of condition (b) alters this result, or affects the proof. In fact, it can be shown that it does not, since the reduction in  \cite{raf2005} involves the construction of an RAF that is automatically closed.

It is also of interest to know how inhibition affects the probability of forming a viable RAF, when $I$ is a random assignment. Notice that inhibition is a much stronger notion than catalysation - since if a reaction is inhibited by just one molecule, then no matter how many molecules might catalyse that reaction, it is prevented from taking place. Thus we might expect that even low rates of inhibition could be a major obstruction to the formation of a viable RAF.   However, we show here that provided the inhibition rate is sufficiently small, Theorem \ref{gentheorem2} still holds.  To state this we first formalize the model by extending (C1) and (C2) to the following three conditions (which reduce to (C1) and (C2) upon setting $\epsilon =0$).

\begin{itemize}
\item[(C1)] The events $\E(x,r)$ that $x$ catalyses reaction $r$, and the events $\F(x,r)$ that $x$ inhibits reaction $r$ are independent across all pairs $(x, r)$ in $X \times \R$ are independent across all pairs $(x,r)$  in $X \times \R$.

\item[(C2)]  As stated previously near the start of Section~\ref{catraf}.

\item[(C3)]  For some constant $\epsilon \geq 0$, the expected number of molecular species that inhibit any given reaction is at most $\epsilon$.
\end{itemize}

Notice that part (a) of Theorem~\ref{gentheorem} applies automatically to the more restrictive notion of an inhibition viable RAF.   However part (b) does not, and here we present a 
stronger result, which implies Theorem~\ref{gentheorem}(b) (upon taking $\epsilon=0$).   The proof of this theorem is presented in the Appendix.

\begin{theorem}
\label{gentheorem2}
Under the extended conditions (C1)--(C3), if $0\leq \epsilon\leq  \exp(-K\overline{c})$, where  $\overline{c} =\overline{\mu}\cdot \frac{|X|}{|R|}$ is the average expected number of molecular species that catalyse each reaction,  then the probability that there exists an RAF for $\Q$   is at least $1 - \psi(\lambda),$  where $\psi(\lambda) = 
 \frac{k(2ke^{-\mu\lambda/K})^t}{1 - 2ke^{-\mu\lambda /K}}  \rightarrow 0$ exponentially fast as
$\lambda \rightarrow \infty$.

\end{theorem}

\subsection{Kinetic RAF framework} \label{krafsection}

Here we extend previous work by introducing the concept of a kinetic CRS, in which every reaction has an associated rate, and all molecules diffuse away into the environment at constant rate. We then define a kinetic RAF, which, informally, is an RAF in which every molecule is produced at least as fast as it is lost (to diffusion, or by consumption in other reactions). This represents the idea that being able to build up a sufficient local concentration of molecules is a necessary condition for RAFs to form.

{\bf Definition:} A {\em kinetic CRS} is a tuple $Q = (X,\R,F,C,v)$ where $X, \R$, $F$ and $C$ are defined in the same way as for a simple CRS, and $v:\R \rightarrow \mathbb{R}_{\geq 0}$ is a {\em rate function}, where for each $r\in \R$, $v(r)$ is the {\em rate} of $r$.

For any subset $\R' \subseteq \R$, The {\em stoichiometric matrix} ${\bf S}_{\R'}$ is the $|{\rm supp}(\R') \setminus F| \times |\R'|$ matrix with rows indexed by the non-food molecule types involved in $\R'$ and columns indexed by the reactions in $\R'$, where ${\bf S}_{ij} \in \mathbb{Z}$ is the net number of molecule type $i$ produced by reaction $j$. The {\em rate vector} ${\bf v}_{\R'}=[v(r_1), v(r_2), \dots, v(r_{|\R'|})]^T$ lists the rates of each reaction in $\R'$. Then, ${\bf S}_{\R'}{\bf v}_{\R'}$ is a vector of the net rates of production of each molecule type in supp$(\R') \setminus F$. Let $\delta \geq 0$ be the {\em diffusion rate}.

A subset $\R' \subseteq \R$ is a {\em kinetic RAF} (kRAF) if and only if the following properties hold (where ${\bf 1}$ is a $|\supp(\R') \setminus F| \times 1$ column vector of $1$s):
\begin{itemize}
\item[(a)] $\R'$ is an RAF for $\Q$;
\item[(b)] $\R'$ is closed;
\item[(c)]  The following inequality holds:  \begin{equation}
{\bf S}_{\R'}{\bf v}_{\R'}-\delta{\bf 1}\geq {\bf 0}.
\label{kraf}
\end{equation}
\end{itemize}

Note that we do not include food molecules in the rows of ${\bf S}_{\R'}$. An RAF $\R'$ is not guaranteed to contain any reactions which generate food molecules, but will necessarily contain at least one reaction with at least one food reactant. In that case, if we were to include the rows corresponding to those food molecules, they would have only negative entries, causing the RAF $\R'$ (which might otherwise satisfy the properties of a kRAF) to formally fail to be a kRAF.

The diffusion rate $\delta$ represents the rate at which molecules diffuse away into the environment. Diffusion is unavoidable in chemical systems, and as molecules diffuse away, their concentrations drop until they are no longer available to sustain local reactions. A CRS occurring in the ocean or a ``pond" might have a larger $\delta$ than one occurring in a hydrothermal vent, which may in turn have a larger $\delta$ than a CRS confined within a lipid membrane \cite{molbiol}.

The idea of searching for kRAFs within a kinetic CRS is related to the idea in chemical organisation theory (COT) of searching for self-sustaining chemical organisations within an algebraic chemistry \cite{cot2007}. The definitions of the stoichiometric matrix coincide, and the qualifying condition \eqref{kraf} for a kRAF is similar to the qualifying condition for an organisation to be self-sustaining \cite{cot2007} (however in COT there is no diffusion term; note that ${\bf S}_{\R'}{\bf v}_{\R'}> 0$ is necessary but not sufficient for a subset $\R' \subseteq \R$ to be a kRAF). Furthermore, in COT the entries of the vector ${\bf v}$ are not fixed - we are free to choose a set of values that makes the system self-sustaining, and indeed the definition of self-sustaining is simply that such a set of values can be found. In contrast, the reactions rates in a kinetic CRS are pre-determined constraints within which we can (in principle) go looking for a subset $\R'$ of reactions that satisfies \eqref{kraf}. While we propose that this set up is more chemically realistic, the following theorem shows that such a search is unlikely to be useful in general. We show that determining whether or not $\R$ contains a kRAF is NP-complete when $\delta=0$  (we expect a similar result applies when $\delta>0$ but our proof, presented in the Appendix, applies to the zero-diffusion case).

\begin{theorem}
Given a kinetic CRS $Q = (X,\R,C,F,v)$ with diffusion rate $\delta=0$, the problem of determining whether or not $\R$ contains a kRAF is NP-complete.
\label{krafsearch}
\end{theorem}

The closely related problem in COT of deciding whether or not an algebraic chemistry contains an organisation is also NP-complete \cite{cot2008}. Although Theorem~\ref{krafsearch} shows that we cannot hope to efficiently find kRAFs within a kinetic CRS, it is easy to check (in polynomial time) whether or not a given RAF is a kRAF, and since RAFs can be found in polynomial time \cite{raf2004}, it may be feasible to discover kRAFs in a kinetic CRS by first ignoring the rate function $v$ and finding a sample of RAFs, then deciding whether or not any are viable under $v$.

One weakness of the kRAF concept is that reaction rates are fixed - in real systems, the rate of a reaction is a function of the concentrations of its reactants, catalysts and inhibitors. Although the concept of concentration currently has no direct meaning in the RAF framework, previous work has used dynamical simulations to study the changes in concentrations of molecules in small RAF sets \cite{raf2012a,raf2012b}.

\section{Concluding comments}

Due to the utility of polymers in modern life, much of the  theoretical and experimental work on the origin of life problem has focussed on system of polymers, and in \cite{raf2005} it was shown that the level of catalysis need only increase linearly as the number of molecules increases, in order to maintain a high probability of RAFs occurring. We have presented a  generalisation of this result, showing that under mild assumptions, the same linear bound applies to a system in which the molecules are not necessarily polymers.  Furthermore, partitioned systems were shown to support the development of RAFs similarly to typical systems containing only one type of polymer, and the effect of the pattern of catalysis on the emergence of RAF sets was explored. Previous research into template-based catalysis \cite{raf2012b} and recent work incorporating more realistic patterns of catalysis \cite{tuebingen} have indicated that the emergence of RAFs is quite robust to the the structure of the underlying reaction system, a conclusion which this paper supports.

This research was performed in an effort to better understand the ``symbiotic coexistence" of peptides and nucleic acids in living organisms, as well as the potential role of this reciprocity in early chemical evolution (as highlighted recently by \cite{lili}).  While the results presented here are a far cry from deep insights revealing fundamental truths about the origin of life, this extension of previous work on chemical reaction systems represents an incremental gain in understanding, which can hopefully contribute to an eventual bigger picture. In particular, this paper supports the experimental work of Carter, et al. \cite{lili} and encourages further experimental work on the topic.

We have also introduced and studied two new concepts in RAF theory: closed RAF sets, and kinetic chemical reaction systems. A closed RAF set is a RAF set in the standard sense, with the additional property that ``every reaction that can occur, does occur". More specifically, this means that if the existing subset of reactions is able to produce all the reactants and at least one catalyst of a reaction outside of the subset, then that reaction should be included the subset. A closed RAF is a subset of reactions that has ``absorbed" every such reaction. 

The kinetic RAF framework was developed in response to criticism levelled at RAF theory for not accounting for the fact that reactions progress at different rates. Kinetics is a fundamental part of real chemistry, so while the strength of RAF theory perhaps lies in its simplicity, the development of a kinetic extension is appropriate. A centerpiece of previous RAF theory investigations has been the search algorithm from \cite{raf2004}, which runs in polynomial time and which has  allowed chemical reaction systems of various sizes and properties to be investigated computationally \cite{raf2004, raf2011}. Therefore, a similar algorithm for detecting kinetically viable RAFs inside a kinetic CRS would be a promising start for the development of a theory of kinetic RAFs. Unfortunately, a reduction from the NP-complete problem 3-SAT showed that detecting a kinetic RAF within a kinetic CRS is unlikely to be productive in general.  However, it is possible to construct RAFs efficiently and for each of these one can test readily whether such an RAF is a kRAF.

\section{Appendix}
\subsection{Proof of Theorem~\ref{gentheorem} and Theorem~\ref{gentheorem2}}

Let $c(r)$ denote the expected number of molecular species that catalyse reaction $r$, let $c_l = \min\{c(r): r \in \R\}$ and $c_u = \max\{c(r): r \in \R\}$ denote the lower and upper bounds on these values, respectively, and let
$\overline{c} = \frac{1}{|\R|} \sum_{r \in \R} c(r)$ denote the average value.   
Now, note that $$\overline{\mu} = \overline{c} \cdot \frac{|\R|}{|X|},$$
and (C2) gives:
$$c_u/c_l \leq K, c_u \leq K \overline{c}, \mbox{ and } c_l \geq \overline{c}/K.$$

Next we establish the following variation on a lemma from \cite{raf2005}. 
\begin{lemma}
Consider a random CRS $\Q = (X, \R, C, I, F)$, satisfying (C1)--(C3).  For a reaction $r$ let $q_r$ be the probability that either no species in $X$ catalyses $r$ or at least one species in $X$ inhibits reaction $r$ .
\begin{itemize}
\item[(i)] $q_r \geq 1-c_u$,
\item[(ii)] $q_r \leq \exp(-c_l) + \epsilon.$
\end{itemize}
\label{genlemma}
\end{lemma}

Let $p(x,r) = \PP(\E(x,r))$ denote the probability that species $x$ catalyses reaction $r$, and let $p'(x,r) = \PP(\F(x,r))$ denote the probability that $x$ inhibits $r$.  
Note that $1-q_r$ is the probability that at least one species in $X$ catalyses $r$ and no species in $X$ inhibits $r$ and so, by  
condition (C1), we have:
\begin{equation}
\label{catin}
1-q_r =  (1- \prod_{x \in X} (1-p(x,r))) \cdot \prod_{x \in X} (1- p'(x,r)). 
\end{equation}
Thus,   $$q_r \geq \prod_{x \in X} (1-p(x,r)) \geq 1- \sum_{x \in X} p(x,r),$$ and
$\sum_{x \in X} p(x,r)$ is the expected number of species that catalyse $r$, which by (C2) is at most $c_u$. Thus, $q_r \geq 1-c_u$ which establishes part (i). 

For part (ii) we have from (\ref{catin}):
 $$q_r \leq \prod_{x \in X} (1-p(x,r))) + (1- \prod_{x \in X} (1- p'(x,r))).$$
and since $$ \prod_{x \in X} (1-p(x,r))) \leq \exp(-\sum_{x \in X}p(x,r)) \leq \exp(-c_l)$$  and $$1- \prod_{x \in X} (1- p'(x,r)) \leq \sum_{x \in X} p'(x,r) \leq \epsilon$$
(by (C2) and (C3)) we obtain the claimed inequality in part (ii). 

To establish Theorem~\ref{gentheorem} part (a), consider the reactions $r \in \R$ such that $\rho(r) \subseteq F$ (we will call these {\em primary} reactions). The number of distinct, potentially reacting pairs of food molecules is 
$${ |F| \choose 2} + |F| = \frac{|F|\left(|F|+1\right)}{2}$$
(this includes pairs of identical molecules). Since $\R$ is finite, there exists $\tau \leq |\R|$ such that no pair of molecules in the system can react in greater than $\tau$ ways to produce $\tau$ distinct products. Then the number of primary reactions in $\R$ is bounded above by $\frac{\tau}{2}\left(|F|^2 +|F|\right)$. By  Lemma~\ref{genlemma} (i), the probability $Q_\R$ that none of the primary reactions are catalysed satisfies:
\begin{equation}
Q_\R \geq (1-c_u)^{\frac{\tau}{2}\left(|F|^2 +|F|\right)}
\label{Q}
\end{equation}
Now the probability that at least one of the primary reactions is catalysed is $1 - Q_\R$, and this is clearly a necessary (but not sufficient) condition for there to be an RAF for $\Q$. It follows that 
$$\PP(\exists \mbox{ RAF for } \Q) \leq 1 - Q_\R \leq 1- (1-c_u)^{\frac{\tau}{2}\left(|F|^2 +|F|\right)}   \leq  1- (1-\overline{c}/K)^{\frac{\tau}{2}\left(|F|^2 +|F|\right)}.$$
Now if $\overline{\mu}  \leq \lambda \cdot \frac{|R|}{|X|}$ then $\lambda \leq \overline{c}$ and so if we take
$$\phi(\lambda) = 1 - (1-\lambda/K)^{\frac{\tau}{2}\left(|F|^2 +|F|\right)}$$
and note that $\phi(\lambda) \rightarrow 0$ as $\lambda  \rightarrow 0$, Theorem~\ref{gentheorem} part (a) now follows.

To establish Theorem~\ref{gentheorem2} (which implies Theorem  \ref{gentheorem}(b)) , let $q_{-} := \exp(-c_l) + \epsilon.$  By the assumption on $\epsilon$ stated in part (b), we have: 
\begin{equation}
\label{qeq2}
q_{-} \leq \exp(-\overline{c}) + \exp(-\overline{c}{K}) \leq 2 \exp(-\overline{c}{K}).
\end{equation}

By Lemma~\ref{genlemma} (ii), for any $s\geq t$ (recalling that $F = \alpha_t$), the probability that a species $x \in X(s+1)$ cannot be produced from reactants in $\alpha_s$ is at most $(q_{-})^{cs}$ (since by (S2) we know that there exist at least $cs$ reactions producing $x$ from reactants in $\alpha_s$, so the only way for $x$ to fail to be produced is if each such reaction has either no catalyst in $X$ or an inhibitor in $X$).
 
Hence $N_s$ denotes the number of species in $X(s+1)$ which cannot be produced by catalysed and uninhibited reactions from reactants in $\alpha_s$. Then the expected value of $N_s$ ($\EE[N_s]$)  is at most $|X(s+1)| (q_{-} )^{\nu s}$, which by (S1) is in turn bounded above by $k^{s+1} (q_{-})^{\nu s}$.  In particular, since $\PP(N_s >0) \leq \EE[N_s]$, the probability  (let us call it $W_{s+1}$) that there exists at least one species in $X(s+1)$ which cannot be produced from reactants in $\alpha_s$ satisfies $$W_{s+1} \leq k^{s+1} (q_{-})^{ \nu s}.$$

Let us say a species in $X$ is {\em problematic} if  each reaction producing that species is either not catalysed by any molecule in $X$ or is inhibited by at least one molecule in $X$. Then the probability that there exists a problematic species in $X$ is $\sum_{s=t}^{m-1} W_{s+1}$, which satisfies
$$\sum_{s=t}^{m-1} W_{s+1} \leq k \sum_{s=t}^{m-1} k^{s} (q_{-})^{\nu s} \leq  k \sum_{s=t}^{m-1} (2k\exp(- \nu \overline{c}/K))^s,$$
where the second inequality applies (\ref{qeq2}). 
Thus the probability that there are no problematic species in $X$ is $1 - \sum_{s=1}^{m-1} W_{s+1}$. A lower bound on this quantity is given by
\begin{align}
1 - \sum_{s=t}^{m-1} W_{s+1} &\geq 1 - k \sum_{s=t}^{m-1} (2k\exp(- r\overline{c}/K))^s \notag \\
&\geq 1 - k \sum_{s=t}^\infty (2k\exp(-\nu \overline{c}/K))^s  \notag \\
&= 1 - \frac{k(2k\exp(-\nu\overline{c}/K))^t}{1 - 2k\exp(-\nu\overline{c}/K)}. \notag
\end{align}
Finally, noting that there being no problematic species in $X$ is a sufficient condition for $\Q$ to have an RAF $\R'$ (indeed one with supp$(\R') = X$), we see that $1 - \sum_{s=1}^{m-1} W_s$ is a lower bound on 
$\PP(\exists \mbox{ RAF for } \Q)$. Hence taking 
$$\psi(\lambda) = \frac{k(2ke^{-\mu\lambda/K})^t}{1 - 2ke^{-\mu\lambda /K}}$$
and noting that $\psi(\lambda ) \rightarrow 0$ as $\lambda  \rightarrow \infty$, part (b) follows (observe that this RAF is closed, since it involves all molecules in $X$). 
This completes the proof.

\subsection{Proof of Theorem~\ref{krafsearch}}

Firstly, given a kinetic CRS $\Q = (X,\R,C,F,v)$ and a subset $\R'$ of $\R$ it can be checked in polynomial time whether $\R'$ is a kRAF for $\Q$, and so 
the question of whether or not $\R$ contains a kRAF is in the complexity class NP.   We will show that this question is NP-complete for the case $\delta=0$  by exhibiting a polynomial-time reduction from the NP-complete problem 3-SAT.  Suppose we have an instance of 3-SAT,  which is an expression $P$ written in conjunctive normal involving a set $Y=\{y_1, \ldots, y_n\}$ of literals, and with each clause consisting of a disjunction of at most three variables (a literal $y_j$ or its negation $\overline{y}_j$).   Thus we can write $P$ in the form 
$$P = C_1 \wedge C_2 \wedge \cdots \wedge C_k\ \mbox{ where } C_i= \vee_{j \in T(i)}y_j \vee_{j \in F(i)} \overline{y}_j.$$

For example, $P = (y_1 \vee \overline{y}_2 \vee y_3) \wedge (y_2 \vee \overline{y_3} \vee \overline{y}_4) \wedge(\overline{y}_1  \vee y_2  \vee \overline{y}_4)$ would be an instance of 3-SAT for 
$Y=\{y_1, y_2, y_3, y_4\}$.

Here $T(i)$ and $F(i)$ are subsets of $\{1, \ldots, n\}$ that describe which elements of $Y$ are in $C_i$ as a literal or a negated
literal (respectively). Since each clause has at most three variables, $|T(i)|+|F(i)| \leq 3$.
We say that $P$ has a {\em satisfying assignment} if there is a function $S: Y \rightarrow \{  \true,\false \}$ so that for each clause $C_i$ in $P$, there exists $j \in T(i)$ for which $f(y_j) = \texttt{true}$ or a $j \in F(i)$ for which $f(y_j) = \false$. In the example above, setting $S(y_1)= \true$, $S(y_2)=S(y_4)=\false$, and $S(y_3)$ to be either \texttt{true} or \texttt{false} provides a satisfying assignment for $P$.

Given $P$ we will construct a catalytic reaction system $(X,\R, C)$, food set $F$, and rate function $v$ so that $\Q_P = (X, \R, C, F, v)$ has a kRAF if and only if $P$ has a satisfying assignment.  

We take $F = \{f_1, \ldots, f_n\}$,  and let $\overline{Y} = \{\overline{y}_1, \ldots, \overline{y}_n\}$. Set 
$$X = F \cup Y \cup \overline{Y} \cup \{y_jT: j=1, \ldots, n\} \cup \{\overline{y}_j T: j =1, \ldots, n\} \cup \{\theta_1, \ldots, \theta_k\} \cup \{\omega, T\}.$$
The reactions, associated rates, and catalysts are described as follows:

\begin{align}
&f_j \rightarrow y_j \mbox{ for each $1\leq j \leq n$; at rate $k+1$; catalysed by $y_jT$}; \label{f-to-y} \\
&f_j \rightarrow \overline{y}_j  \mbox{ for each $1\leq j \leq n$; at rate $k+1$; catalysed by $\overline{y}_jT$}; \label{f-to-y-neg} \\
&y_j + T \rightarrow y_jT  \mbox{ for each $1\leq j \leq n$; at rate $0<\epsilon<1/n$; catalysed by $T$}; \label{y-to-cat} \\
&\overline{y}_j + T \rightarrow \overline{y}_jT  \mbox{ for each $1\leq j\leq n$; at rate $0<\epsilon<1/n$; catalysed by $T$}; \label{y-neg-to-cat} \\
&y_j \rightarrow \theta_i \mbox{ for each pair $(i,j)$ with $j \in T(i)$; at rate 1; catalysed by $T$};  \label{y-to-theta} \\
&\overline{ y}_j \rightarrow \theta_i \mbox{ for each pair $(i,j)$ with $j \in F(i)$; at rate 1; catalysed by $T$}; \label{y-neg-to-theta} \\
&\theta_1+ \cdots + \theta_k \rightarrow T\mbox { at rate 1; catalysed by $T$}; \label{T-gen} \\
&y_j + \overline{y}_j \rightarrow \omega \mbox{ for each $1\leq j \leq n$,  at rate $k+2$; catalysed by $T$}. \label{annihilate}
\end{align}

First suppose that  $\Q_P$ contains a kRAF $\R'$; we will show that $P$ has a satisfying assignment.  Since $\R'$ is an RAF, it is non-empty. Therefore, the molecule $T$ must be produced, since every reaction in $\R$ is catalysed by either $T$ or some molecule that is produced from $T$.  This in turn requires that for each $1\leq i \leq k$, $\theta_i$ is produced, and therefore, for each $1\leq i \leq k$ there exists $j \in T(i)$ such that $y_j$ is produced or $j \in F(i)$ such that $\overline{y}_j$ is produced. Furthermore, for each value of $1\leq j \leq n$, at most one of the molecules $y_j, \overline{y}_j$ is produced, since otherwise by the closure property the $j$th reaction described by \eqref{annihilate} would be contained in $\R'$, which would destroy both $y_j$ and $\overline{y}_j$ faster than either is produced and violate the rate property of the kRAF $\R'$. A satisfying assignment $S$ for $P$ is now provided by setting $S(y_j)$ to be $\true$ (respectively \false) if $y_j$ is produced by some reaction  in $\R'$ (respectively not produced by some reaction in $\R'$). Note that $S$ is a satisfying assignment even in the case where neither of $y_j, \overline{y}_j$ is produced for some $j\in\{1,\dots,n\}$, since in that case $S(y_j)$ can be chosen arbitrarily. 

Conversely suppose that $P$ has a satisfying assignment $S$; we will show that $\Q_P$ contains a kRAF. 
Let $\R'$ consist of reaction \eqref{T-gen} together with the following reactions:
\begin{itemize}
\item for each $j\in\{1,\dots,n\}$ such that $S(y_j)=\texttt{true}$, include the $j$th reaction from \eqref{f-to-y}, the $j$th reaction from \eqref{y-to-cat}, and every reaction from \eqref{y-to-theta} such that $j \in T(i)$;
\item for each $j\in\{1,\dots,n\}$ such that $S(y_j)=\texttt{false}$, include the $j$th reaction from \eqref{f-to-y-neg}, the $j$th reaction from \eqref{y-neg-to-cat}, and every reaction from \eqref{y-neg-to-theta} such that $j \in T(i)$.
\end{itemize}
To show that $\R'$ is a kRAF, we must show that it is a closed RAF which satisfies the rate requirement (Equation~\ref{kraf}). It is easy to see that
\begin{align}
\rho(\R') &= \c_{\R'}(F) \notag \\
&=F \cup \{ y_j : S(y_j) = \true\}\cup \{ \overline{y}_j : S(y_j) = \false\} ~\cup \label{closure} \\
&~ \{  y_jT : S(y_j)=\true \} \cup \{\overline{y}_jT : S(y_j) = \false     \} \cup \{\theta_1, \dots, \theta_k\}\cup \{T\}, \notag
\end{align}
so $\R'$ is $F$-generated. Moreover, every reaction is catalysed by exactly one molecule from the set $\{T\} \cup \{ y_jT : S(y_j) = \true\} \cup \{ \overline{y}_jT : S(y_j) = \false\}$, and since this union is a subset of $\c_{\R'}(F)$, $\R'$ is also reflexively autocatalytic and is therefore an RAF set. 

$\R'$ is closed if there are no reactions $r \in \R\setminus \R'$ such that there exists $(x,r)\in C$ with $\{x\} \cup \rho(r) \subseteq \c_{\R'}(F)$. By the construction of $\R'$, $\R \setminus \R'$ contains the following reactions:

\begin{itemize}
\item $f_j \rightarrow y_j  \mbox{  for each $j$ such that $S(y_j) =$ \false}~ \mbox{(catalysed by $y_jT$)};$
\item  $f_j \rightarrow \overline{y}_j \mbox{  for each $j$ such that $S(\overline{y}_j) =$ \true}~ \mbox{(catalysed by $\overline{y}_jT$)};$
\end{itemize}
(the catalysts of these reactions are not contained in $\c_{\R'}(F)$)
\begin{itemize}
\item $y_j + T  \rightarrow y_jT   \mbox{  for each $j$ such that $S(y_j) =$ \false}$;
\item $\overline{y}_j + T \rightarrow \overline{y}_jT \mbox{  for each $j$ such that $S(\overline{y}_j) =$ \true}$;
\item $y_j \rightarrow \theta_i \mbox{ for each pair $(i,j)$ with $j \in T(i)$ and $S(y_j)=$ \false}$;
\item $\overline{y}_j \rightarrow \theta_i \mbox{ for each pair $(i,j)$ with $j \in F(i)$ and $S(y_j)=$ \true}$;
\end{itemize}
(Other than $T$, the reactants of these reactions are not contained in $\c_{\R'}(F)$)
\begin{itemize}
\item $y_j + \overline{y}_j \rightarrow \omega \mbox{ for each $j \in \{1, \dots, k\}$}$.
\end{itemize}
For each value of $j$, exactly one of the two reactants of this last reaction is contained in $\c_{\R'}(F)$. Hence, $\R'$ is closed.

It remains to show that $\R'$ satisfies the rate condition from section \ref{krafsection}. Recall that the rows of the stoichiometric matrix are indexed by the molecules in the set 
$$\supp(\R')\setminus F = \c_{\R'}(F)\setminus F,$$
the elements of which are given by \eqref{closure}. 

The molecules $\{y_j : S(y_j) = \true\}$ are each produced at rate $k+1$ from $f_j$, used up at rate $\epsilon>0$ to produce $y_jT$, and used up at rate $1$ by each of the reactions $\{ y_j \rightarrow \theta_i:  j \in F(i)\}$. Since there are $k$ clauses, there are at most $k$ values of $i$ for which $j \in T(i)$. Hence the overall rate of production of each molecule $y_j$ is at least $k+1-(k+\epsilon) = 1-\epsilon>0$, which satisfies the rate condition. A similar argument can be made to show that the molecules $\{ \overline{y}_j : S(\overline{y}_j) = \false\}$ also satisfy the condition.

The molecule $T$ is produced at rate $1$ by the reaction $\theta_1 + \dots + \theta_k \rightarrow T$, and used up at rate $0< \epsilon<1/n$ by each of the $n$ reactions forming $y_jT$ or $\overline{y}_jT$. Hence the overall rate of production of $T$ is guaranteed to be positive.

Consider the molecules $\theta_1,\dots,\theta_k$. $\theta_i$ is produced at rate $1$ by each reaction from \eqref{y-to-theta} or \eqref{y-neg-to-theta} that is included in $\R'$, of which there are at least one (since $P$ has a satisfying assignment). $\theta_i$ is also used up at rate $1$ by reaction \eqref{T-gen}, hence the overall rate of formation of $\theta_i$ is non-negative. 

Finally, noting that the molecules $y_jT$ and $\overline{y}_jT$ are all produced at rate $\epsilon>0$ and are not used by any reaction, we see that every molecule in $\supp(\R')\setminus F$ is produced at least as fast as it is used up. This shows that $\R'$ is a kRAF, and so completes the reduction.


\begin{backmatter}

\section*{Competing interests}
The authors declare that they have no competing interests.

\section*{Author's contributions}
J.S. formulated the partitioned model and performed the simulations and analysis; W.H. wrote the code; M.S. formulated the closed, kinetic and inhibitory extensions; all authors contributed towards writing the paper.

\section*{Acknowledgements}
We thank the University of Canterbury's  BlueFern supercomputing unit for technical support and the use of the IBM Power775 cluster on which the partitioned model simulations were performed. We also thank the Allan Wilson Centre for supporting this research. 

\section*{Endnotes}
a: Peptide nucleic acid (PNA) does exist, however this polymer has a backbone of N-(2-aminoethyl)glycine (AEG) monomers linked by peptide bonds, with nucleobases attached to each monomer, rather than being composed of both nucleotide and amino acid monomers. Interestingly, the recent discovery of AEG production in diverse taxa of cyanobacteria may suggest an information-carrying role for PNA in early life \cite{ban}.

b: tRNA aminoacylation or ``charging" involves the esterification of an amino acid monomer to the relevant tRNA, prior to translation at the ribosome. This is of course an example of a reaction which combines molecules from both ``independent" sets.

\bibliographystyle{bmc-mathphys} 
\bibliography{partbin}      



\newpage
\section*{Figure captions}
\subsection*{Figure 1 - A chemical reaction system}
A simple CRS within the binary polymer model, where the food set consists of all monomers and dimers. The subset $\{r_1, r_2, r_3, r_5 \}$ is the maximal RAF subset (the {\em maxRAF}), while $\{r_1, r_2\}$, $\{r_3 \}$ and $\{r_1, r_2, r_3\}$ are smaller RAFs within the maximal RAF known as {\em subRAFs}. $\{r_1, r_2\}$ is an example of an {\em irreducible} RAF, since no proper subset of it is an RAF. A reaction such as $\{r_3\}$, which consists of a single autocatalytic reaction with food molecules as reactants, is sometimes called a `trivial RAF'.
\label{CRSexample}

\vspace{1cm}
\includegraphics[scale=1.0]{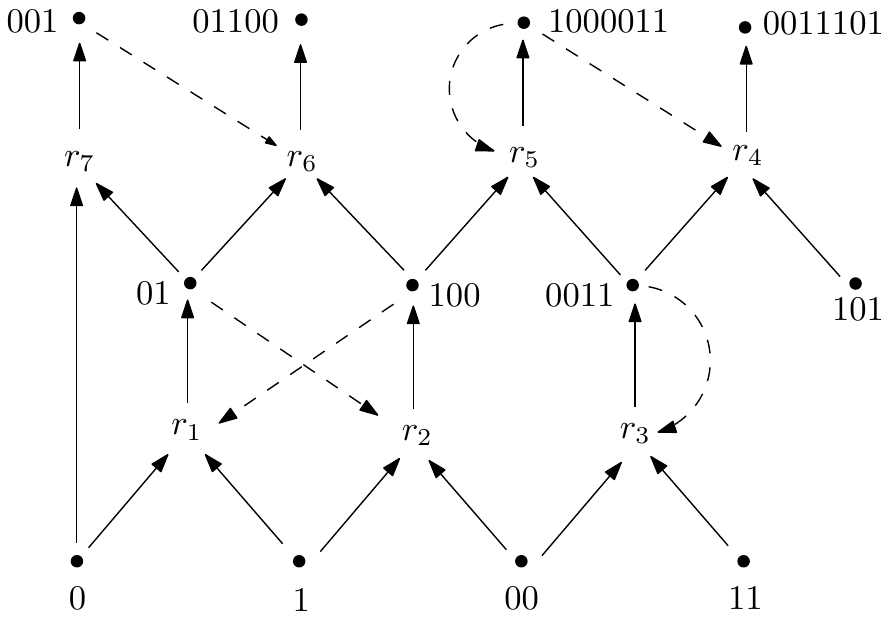}

\subsection*{Figure 2 - Reciprocity of peptides and nucleic acids}
Schematic depicting the mutual catalytic dependence between nucleic acids and peptides in living systems, where a dashed arrow from X to Y indicates that there exist reactions involving molecules in Y which are catalysed by molecules in X. While all possible such arrows are present in the diagram, both groups of molecules are ``closed" in the sense that there are no reactions combining nucleotide and amino acid monomers in the same polymer.
\label{schematic}

\vspace{1cm}
\includegraphics[scale=1.0]{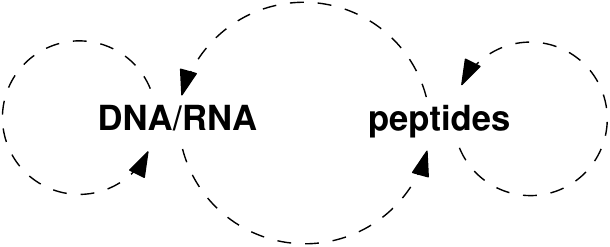}

\subsection*{Figure 3 - A partitioned CRS}
A partitioned CRS within the binary polymer model. One set of molecules is built from the `square' and `circle' monomers, and the other is built from the `triangle' and `hexagon' monomers. The molecules at the bottom of the image comprise the food set and give rise to the other molecules via the ligation and cleavage reactions $r_1, \dots, R_4$. Dashed arrows indicate which molecules catalyse which reactions. In this case, the entire CRS is an RAF. Note that while there is intra- and inter-partition catalysis, there are no reactions involving molecules from both partitions. This is emphasized by the enclosure of each partition within a large circle: of course, in real systems, the molecules would be free to mingle.
\label{partCRS}

\vspace{1cm}
\includegraphics[scale=1.0]{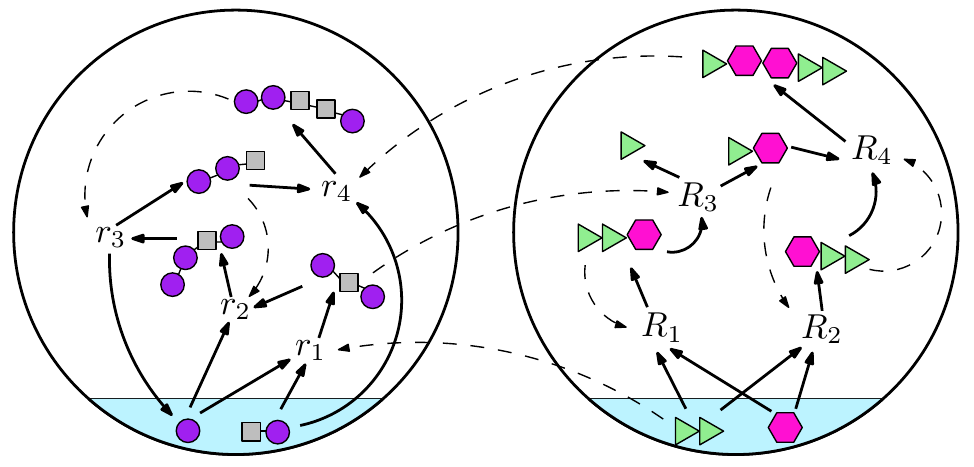}

\subsection*{Figure 4 - Selfish, uniform and reciprocal models}
Visual representations of the three systems investigated via simulations: reciprocal (left), selfish (centre) and uniform (right). Labelled arrows from $X_i$ to $X_j$ indicate that molecules in $X_i$ catalyse reactions involving molecules from $X_j$, with some non-zero probability (given by the label). The catalysis matrix ${\bf P}$ is shown for each system. It can readily be shown that the level of catalysis (expected number of reactions catalysed per molecule) is the same in all three scenarios.
\label{CRSs}

\vspace{1cm}
\includegraphics[scale=0.8]{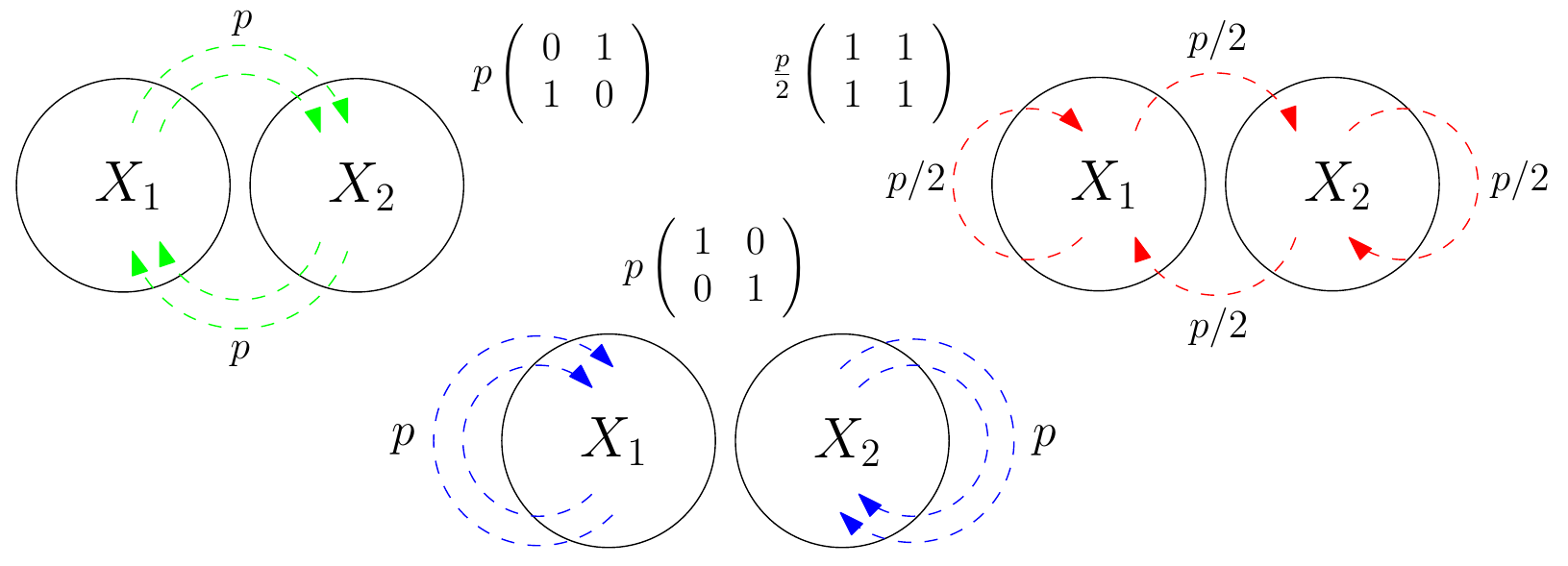}

\subsection*{Figure 5 - Emergence of RAFs in a partitioned system with $n=10$}
      Plot showing how the proportion of CRSs containing an RAF depends on the level of catalysis, for each of the three models. The maximum length of polymers ($n$) is $10$ and the food set consists of all monomers and dimers. The fraction of CRSs containing an RAF is from 500 instances of the model.
\label{n10frac}

\vspace{1cm}
\includegraphics[scale=0.8]{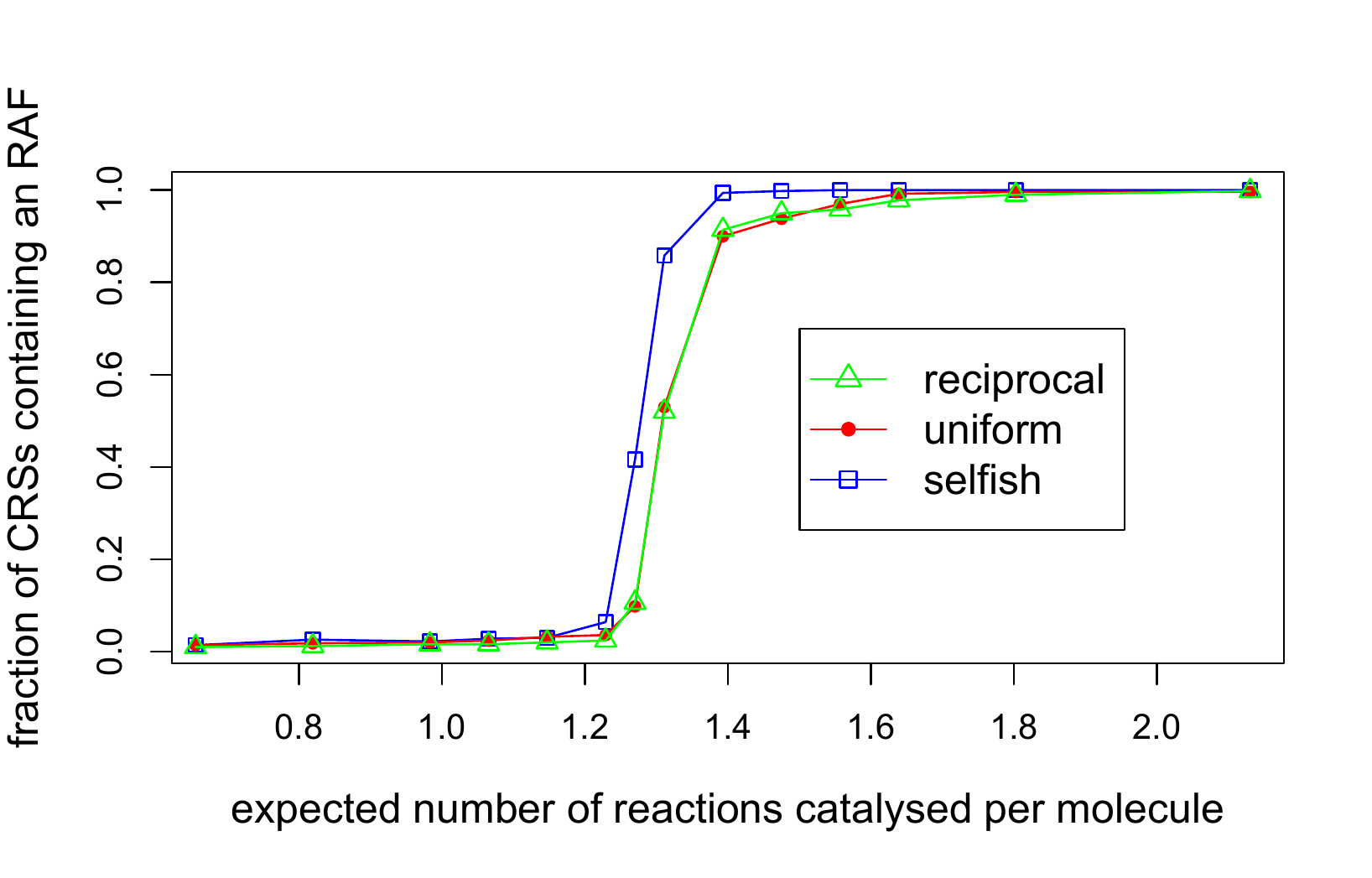}

\subsection*{Figure 6 - Size of the maxRAF in a partitioned system with $n=10$}
Plots showing how the average number of reactions and number of molecules in the maxRAF change as the level of catalysis moves through the transition range, expressed as a proportion of the total number of reactions and molecules in the CRS. Averages are taken over 500 instances of each model where $n=10$ and the food set consists of all monomers and dimers. Instances containing no maxRAF were excluded from the calculation of the average, hence the data points that appear close to zero indicate a small but non-zero average size.
\label{n10size}

\vspace{1cm}
\includegraphics[scale=0.5]{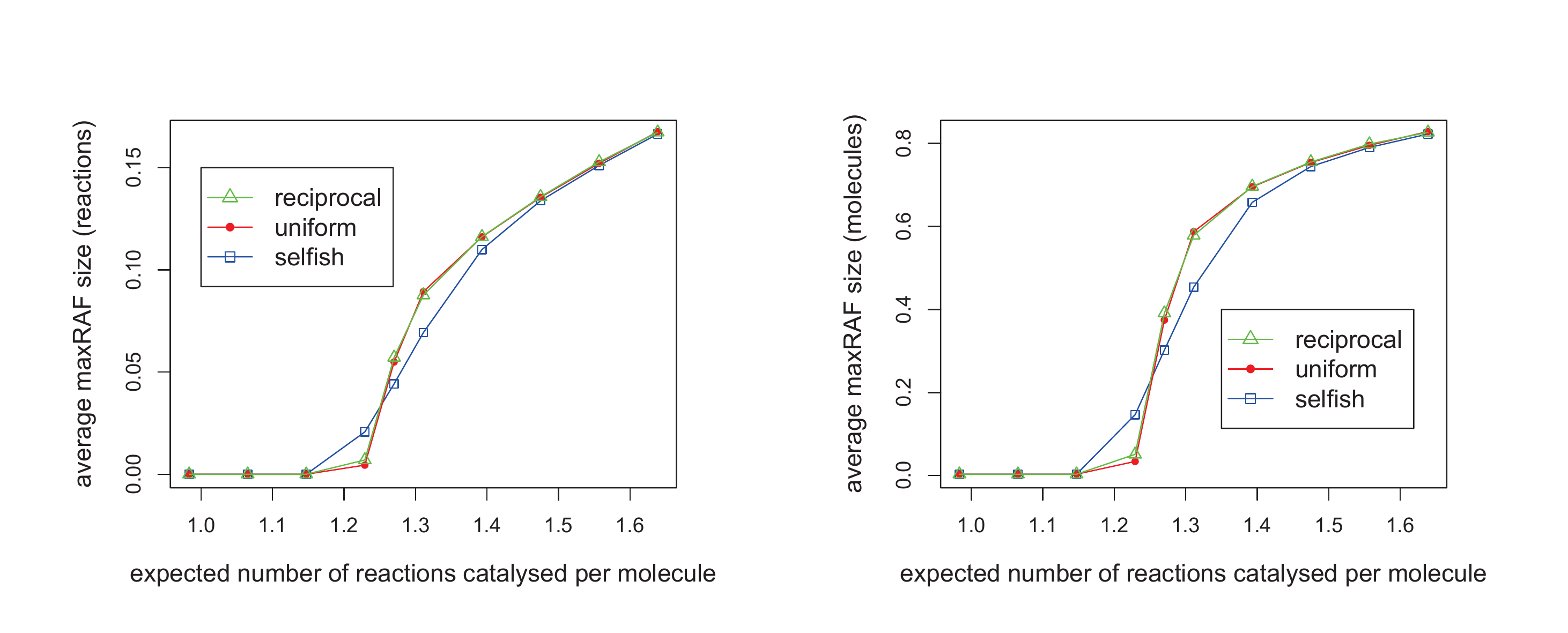}

\subsection*{Figure 7 - Emergence of RAFs in a partitioned system with $n=15$}
Plot showing how the proportion of CRSs containing an RAF depends on the level of catalysis, when the maximum length of polymers in the system ($n$) is 15. The fraction of CRSs containing an RAF is from at least 120 instances of the model.
\label{n15frac}

\vspace{1cm}
\includegraphics[scale=0.8]{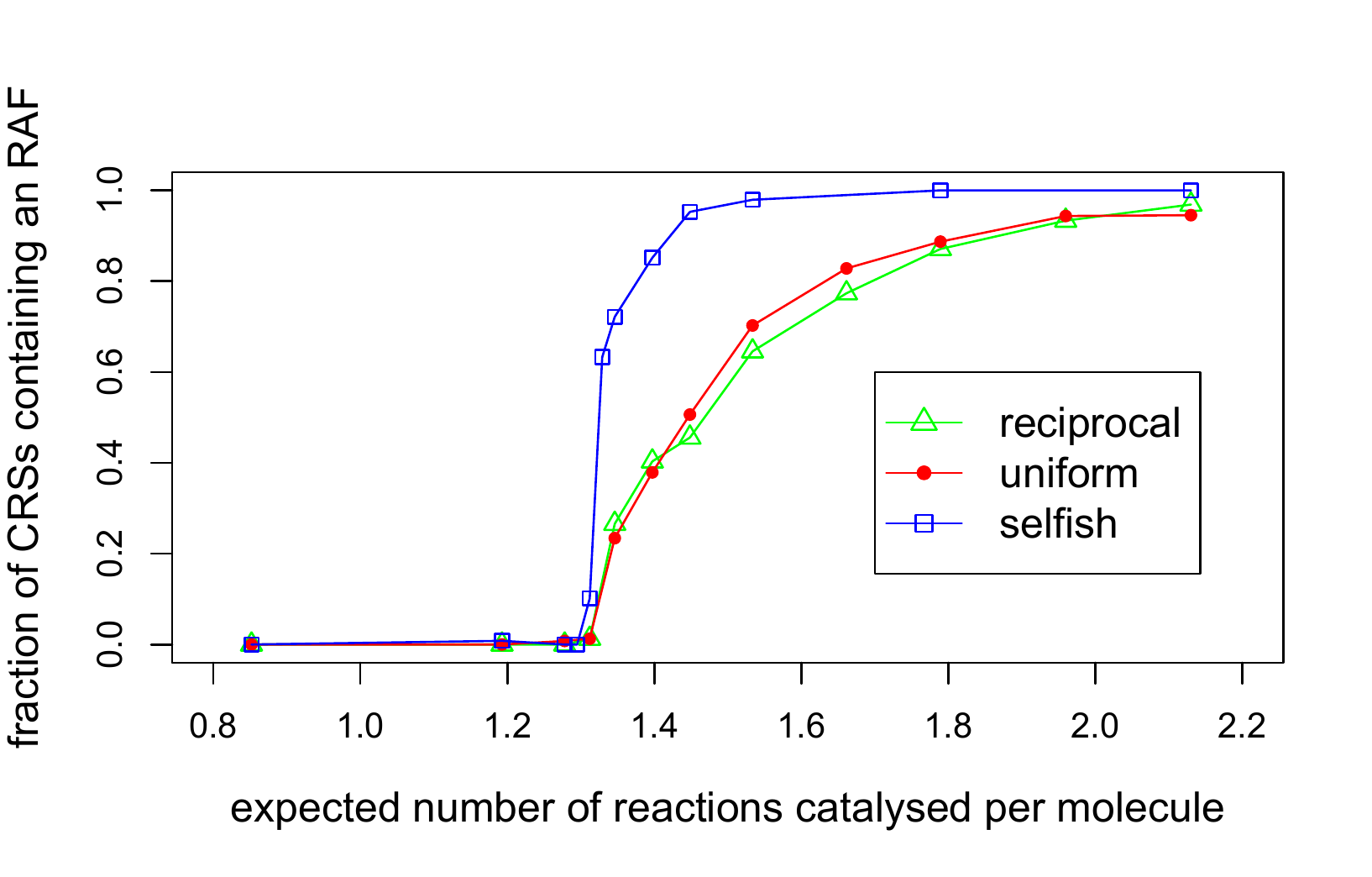}

\subsection*{Figure 8 - Size of the maxRAF in a partitioned system with $n=15$}
Plots showing how the average number of reactions and average number of molecules in the maxRAF change as the level of catalysis is increased, expressed as a proportion of the total number of reactions and molecules in the CRS when $n=15$. Instances containing no maxRAF were excluded from the calculation of the average.
\label{n15size}

\vspace{1cm}
\includegraphics[scale=0.5]{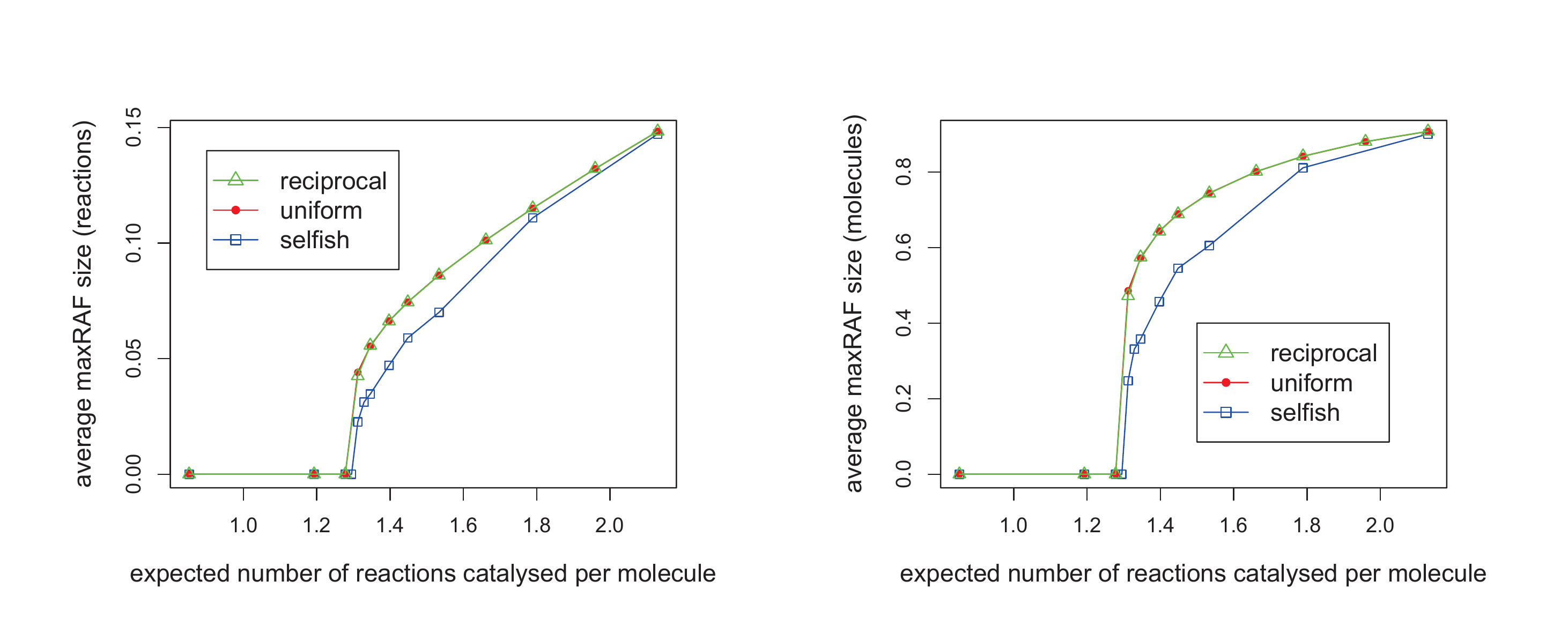}

\end{backmatter}
\end{document}